\documentclass[twocolumn]{aastex61}

\usepackage{threeparttable}
\usepackage{amsmath}

\newcommand*{\rom}[1]{\expandafter\@slowromancap\romannumeral #1@}

\newcommand{\msun}{M$_{\odot}$ }
\newcommand{\myr}{M$_\odot$~yr$^{-1}$ }

\newcommand{\ha}{H$\alpha$ }
\newcommand{\hb}{H$\beta$ }
\newcommand{\nii}{[N{\sc II}] }
\newcommand{\oiii}{[O{\sc III}] }

\newcommand{\sii}{[S{\sc II}] }

\newcommand{\loghn}{log([N{\sc II}]/H$\alpha$) }
\newcommand{\logohb}{log([O{\sc III}]/\hb) }

\newcommand{\kms}{km\,s$^{-1}$ } 

\newcommand{\ferg}{erg s$^{-1}$ cm$^{-2}$ }

\newcommand{\ergs}{erg s$^{-1}$ }

\newcommand{\eden}{cm$^{-3}$ }
\newcommand{\vsigmastellar}{$\sigma_{*}$ }
\newcommand{\vsigmagas}{$\sigma_{gas}$ }

\newcommand{\msigma}{$M_{\bullet}-\sigma~$}
\newcommand{\mstellar}{$M_{\bullet}-M_{*}~$}

\submitjournal{ApJ}

\accepted{January 15, 2021}

\shorttitle{Kinematics and Energetics of Ionization Gas}
\shortauthors{Vayner et al.}

\begin{document}

\title{A Spatially-Resolved Survey of Distant Quasar Host Galaxies:\\ II. Photoionization and Kinematics of the ISM}

\correspondingauthor{Andrey Vayner}
\email{avayner1@jhu.edu}

\author[0000-0002-0710-3729]{Andrey Vayner}
\affiliation{Department of Physics, University of California San Diego, 
9500 Gilman Drive 
La Jolla, CA 92093 USA}
\affiliation{Center for Astrophysics \& Space Sciences, University of California San Diego, 9500 Gilman Drive La Jolla, CA 92093 USA}
\affiliation{Department of Physics and Astronomy, Johns Hopkins University, Bloomberg Center, 3400 N. Charles St., Baltimore, MD 21218, USA}

\author[0000-0003-1034-8054]{Shelley A. Wright}
\affiliation{Department of Physics, University of California San Diego, 
9500 Gilman Drive 
La Jolla, CA 92093 USA}
\affiliation{Center for Astrophysics \& Space Sciences, University of California San Diego,
9500 Gilman Drive 
La Jolla, CA 92093 USA}

\author{Norman Murray}
\affiliation{Canadian Institute for Theoretical Astrophysics, University of Toronto, 60 St. George Street, Toronto, ON M5S 3H8, Canada}
\affiliation{Canada Research Chair in Theoretical Astrophysics}

\author[0000-0003-3498-2973]{Lee Armus}
\affiliation{Spitzer Science Center, California Institute of Technology, 1200 E. California Blvd., Pasadena, CA 91125 USA}


\author[0000-0003-0439-7634]{Anna Boehle}
\affiliation{ETH Zürich Wolfgang-Pauli-Str. 27 8093 Zürich, Switzerland}

\author[0000-0002-2248-6107]{Maren Cosens}
\affiliation{Department of Physics, University of California San Diego, 
9500 Gilman Drive 
La Jolla, CA 92093 USA}
\affiliation{Center for Astrophysics \& Space Sciences, University of California San Diego, 9500 Gilman Drive La Jolla, CA 92093 USA}

\author[0000-0001-7687-3965]{James E. Larkin}
\affiliation{Department of Physics and Astronomy, University of California, Los Angeles, CA 90095 USA}

\author[0000-0001-7127-5990]{Etsuko Mieda}
\affiliation{National Astronomical Observatory of Japan, Subaru Telescope, National Institutes of Natural Sciences, Hilo, HI 96720, USA}

\author[0000-0002-6313-6808]{Gregory Walth}
\affiliation{Observatories of the Carnegie Institution for Science
813 Santa Barbara Street
Pasadena, CA 91101
USA}

\begin{abstract}
We present detailed observations of photoionization conditions and galaxy kinematics in eleven z$=1.39-2.59$ radio-loud quasar host galaxies. Data was taken with OSIRIS integral field spectrograph (IFS) and the adaptive optics system at the W.M. Keck Observatory that targeted nebular emission lines (\hb,\oiii,\ha,\nii) redshifted into the near-infrared (1-2.4 \micron). We detect extended ionized emission on scales ranging from 1-30 kpc photoionized by stars, shocks, and active galactic nuclei (AGN). Spatially resolved emission-line ratios indicate that our systems reside off the star formation and AGN-mixing sequence on the Baldwin, Phillips $\&$ Terlevich (BPT) diagram at low redshift. The dominant cause of the difference between line ratios of low redshift galaxies and our sample is due to lower gas-phase metallicities, which are 2-5$\times$ less compared to galaxies with AGN in the nearby Universe. Using gas velocity dispersion as a proxy to stellar velocity dispersion and dynamical mass measurement through inclined disk modeling we find that the quasar host galaxies are under-massive relative to their central supermassive black hole (SMBH) mass, with all systems residing off the local scaling (\msigma,\mstellar) relationship. These quasar host galaxies require substantial growth, up to an order of magnitude in stellar mass, to grow into present-day massive elliptical galaxies. Combining these results with part I of our sample paper \citep{vayner19a} we find evidence for winds capable of causing feedback before the AGN host galaxies land on the local scaling relation between black hole and galaxy stellar mass, and before the enrichment of the ISM to a level observed in local galaxies with AGN.

\end{abstract}

\section{Introduction} \label{sec:chap3intro}

Today, feedback from supermassive black holes (SMBH) is an integral part of galaxy evolution models. It is commonly used to explain the lack of observed baryons in local massive galaxies \citep{Behroozi10}, the enrichment of the circumgalactic medium with metals \citep{Prochaska14} and the observed local scaling relation between the mass of the galaxy/bulge and the SMBH \citep{Ferrarese00,Gebhardt00,McConnell13}.

The latest observational and theoretical results point to a critical question; at what points does the AGN drive an outflow powerful enough to clear the galaxy of its gas into the surrounding CGM? \citep{King15} According to theoretical work, this typically happens once the galaxy reaches the \msigma relationship \citep{Zubovas14}. However, there has been growing evidence for galaxies with massive SMBH and powerful outflows that are offset from the local scaling relationship \citep{Vayner17}. The origin and evolution of the local scaling relationships with redshift have been an active debate topic over the last decade. When are the local scaling relations established? Are the local scaling relationships the end product of galaxy evolution? Meaning, as galaxies form and evolve, do they fall in and out of the relationships due to rapid growth or feedback process? Do galaxies eventually end up on the local scaling relations once the galaxy or SMBH catch up and finish growing \citep{Volonteri12}? Alternatively, is there an inherent evolution in the scaling relationship with redshift and a symbiosis between the galaxy and SMBH growth? (i.e., evolution in slope, offset, and scatter). Finally, there is still an open question regarding the role of quasar feedback in establishing the relationship and whether the merging of galaxies can produce the \msigma relation following the central limit theorem \citep{Jahnke11}. From a sample of AGN in the COSMOS field \citep{Merloni10} finds an offset in the local scaling relationship between redshift 0 and 2. These authors use SED decomposition with numerous spectral bands to measure the stellar mass of the AGN host galaxy in the redshift range of $1<\rm z<2.2$. From a sample of lensed quasars at $1<z<4.5$ and broadband HST imaging, \cite{Peng06} finds an offset in the local scaling relationship. While \cite{Sun15} using multi-band SED fitting of galaxies in the COSMOS field finds that z$\sim0.2-2$ galaxies are consistent with being on the local scaling relationship. \cite{Schramm13} using HST imaging in the Chandra Extended Deep Field also finds that galaxies at z$\sim0.6-1$ are also consistent with being on the local scaling relationship. In the nearby Universe, there is tentative evidence that all of the most massive black holes ($>10^{9}$ \msun) are systematically more massive relative to their host galaxies \citep{MartinNovarro18}. Fields such as COSMOS or the Extended Chandra Deep Field-South are relatively small in the sky; hence, the number of luminous quasars with massive SMBH is small. Studies that explored the evolution of the local scaling relationships have generally focused on lower-mass black holes with masses $<10^{9}$ \msun. Furthermore, a large fraction of these studies used broadband HST imaging to study the host galaxies of their quasars/AGN. It is often difficult to disentangle the bright AGN emission from the host galaxy at smaller angular separations ($<$0.5\arcsec). These studies have a limited number of filters to measure the stellar population's age and the mass to light ratio. Alternatively, mm-interferometry observations have become an essential tool in measuring the dynamical masses of quasar host galaxies across different redshift ranges. At the highest redshifts (z$>4$), the [CII] 158\micron~line has been the most commonly used tracer of the dynamics of the ISM. There is growing evidence that the most massive ($>10^{9}$ \msun) SMBH in the highest redshift quasars known to date (z$>6$) appear to be over massive for the mass of their host galaxies \citep{Wang13, Venemans16, Decarli18}, indicating that the most massive SMBHs residing in high redshift quasars grow first relative to their host galaxies. At more intermediate redshifts 1$<$z$<$3, some systems also appear to have overly massive SMBH relative to their stellar/dynamical mass \citep{Shields06, Trakhtenbrot15, Vayner17}. While a significant fraction of galaxies with lower SMBH $<10^{9}$ \msun appear closer or within the scatter of the local scaling relations, galaxies with the luminous quasars and massive SMBH appear to be under massive relative to the mass of their SMBH. As outlined by \citep{Lauer07,Schulze14}, the offset from the local scaling relations for the systems with more massive black holes is biased due to the steep decline in the galaxy mass function at the massive end. 

Integral field spectroscopy (IFS) behind adaptive optics is another method with which it is possible to disentangle the bright quasar emission from the extended emission of the host galaxy. A point spread function can be constructed using data channels confined to the broad emission line of the quasar. After the point spread function is normalized, it is subtracted from the rest of the data channels in the cube. This technique was first shown to be able to resolve host galaxies of low redshift ($z<0.2$) luminous type-1 quasars in seeing limited observations \citep{Jahnke04} and extended Ly$\alpha$ emission around high redshift quasars \citep{Christensen06}. Later, when the first near-infrared IFS came online along with their own adaptive optics system, this technique was expanded to samples of higher redshift quasars in search for their host galaxies \citep{Inskip11, Vayner16} and has shown to work on all the 8-10m class near-infrared IFS (e.g., SINFONI, NIFS, and OSIRIS). This technique has shown continued success in seeing limited optical IFS data as well \citep{Herenz15,Arrigoni-Battaia19}. This PSF subtraction routine provides better contrast at smaller angular separations than HST, with an inner working angle of 0.1-0.2\arcsec, compared to $\sim$ 0.5\arcsec for HST \citep{Vayner16}. Although today's near-infrared IFSs are not sensitive enough to detect the stellar continuum from the quasar/AGN host galaxies, they can still detect extended ionized emission, enabling us to extract the dynamical properties of the galaxy \citep{Inskip11, Vayner17} and compare systems to the local scaling relation. However, today, the largest fraction of quasar host galaxy masses still come from HST and mm-interferometric observations. Most likely, selection effects play an important role in determining whether there is a systematic offset from the local scaling relations among the different studies.

Besides measuring the host galaxies and SMBH masses, there are vital open questions regarding the gas phase properties. Galaxies exhibit a correlation between the stellar mass and metallicity across a wide redshift range \citep{Erb06c,Sanders15}. It is often difficult to place galaxies with bright AGN on the mass-metallicity relationship due to limited contrast and the fact that the AGN has a strong impact on the ISM's ionization state. What are the metallicities of the gas in quasar hosts? How does the metallicity in quasar host galaxies evolve with redshift? What is the dominant source of ionization in quasar hosts? What are the star formation rates? One of the best ways to measure the ionization properties of the gas in galaxies is through the BPT (Baldwin,  Phillips  $\&$  Terlevich) diagram \citep{Baldwin81, Veilleux87}. The traditional BPT diagram plots the ratio of \logohb vs. \loghn~and contains two clearly defined sequences: the star-forming sequence and the mixing sequence. The star-forming sequence provides information about the metallicity of HII regions, the stellar ionizing radiation field as well as information on the gas condition in star-forming regions. On the other hand, the mixing sequence consists of gas photoionized by hot stars, AGN, and shocks. It can potentially provide information on the hardness of the AGN ionizing radiation and the metallicity of the gas photoionized by the quasar/AGN \citep{Groves06} and shocks.
Studies of high redshift star-forming galaxies have shown evidence for elevated line ratios relative to low redshift galaxies. At z$\sim$2, the observed elevated line ratios have been attributed to denser ISM conditions \citep{Sanders16} and harder ionizing radiation fields at fixed N/O and O/H abundances relative to typical z=0 galaxies \citep{Strom17}. Evolutionary BPT models by \cite{Kewley13a} are consistent with these observations. The evolutionary BPT models also provide a prediction on the evolution of the mixing sequence between z=0 and 3. The location of the mixing sequence moves to lower \loghn~value at a relatively fixed \logohb value, primarily due to lower on average gas-phase metallicity at higher redshift \citep{Groves06,Kewley13a}. There is tentative evidence that gas photoionized by AGN is consistent with this picture, as there are several galaxies with AGN, which have emission line ratios offset from the local mixing sequence \citep{Juneau14, Coil15, Strom17, Nesvadba17B, Law18}. Given the presence of the AGN, young stars and shocks in quasar host galaxies, it is crucial to spatially resolve the quasar host galaxy to understand the various contributions to gas ionization. In the distant Universe, this generally requires observations with an IFS and adaptive optics. Resolved BPT diagnostics in both nearby and distant AGN/quasar host galaxies have found regions with distinct photoionization mechanisms \citep{Davies14, Williams17, Vayner17}. The question remains whether the ISM condition in the most luminous high redshift quasar host galaxies is different from local AGN and where they lie relative to the mass metallicity relationship.

We have begun a survey to study the host galaxies of z$=1.4-2.6$ radio-loud quasars, which are likely to evolve into the most massive elliptical galaxies in the nearby Universe. The sample consists of eleven objects, selected to have young-jets with sizes up to 20 kpc in order to study their impact on galactic scales at early times. The observations consist of near-infrared IFS observation behind laser-guide-star adaptive optics (LGS-AO) at the W.M. Keck Observatory with the OSIRIS instrument. The survey aims to understand the gas phase conditions and ionization mechanisms in high redshift quasar host galaxies and search for evidence of quasar feedback and weighing the masses of the quasar hosts. The observations target nebular emission lines (\hb,\oiii,\ha,\nii,\sii) redshifted in the near-infrared bands ($1-2.4$ \micron), at the distance of our sample, the angular resolution of the OSIRIS/LGS-AO mode corresponds to approximately 1.4 kpc in projection. 

This paper is part two of two papers focusing on understanding the photoionization mechanisms of gas in radio-loud quasar host galaxies and weigh the mass of the galaxy and SMBH to compare them to the local scaling relations. Refer to paper I \citep{vayner19a} for details on the sample selection, properties, and data reduction. Details on archival HST imaging data set are presented in \S \ref{sec:archival-HST}. Blackhole masses are presented in \S \ref{sec:bh_masses}, we describe how we identify spatially-resolved dynamically quiescent regions in each quasar host galaxy in \S \ref{sec:regions}, resolved BPT diagrams and our interpretation of the line ratios are present in \S \ref{sec:BPT}, dynamical masses of the quasar host galaxies and their place relative to the local scaling relations is presented in \S \ref{sec:dyn-mass} \& \S \ref{sec:galaxy_dyn_mass_sigma}, we discuss our results in broader context of massive galaxy evolution in \S \ref{sec:chapter4discussion} and present our conclusions in \S \ref{sec:conclusions}. Notes on individual sources are presented in \S \ref{sec:appendix}. Throughout the paper we assume a $\Lambda$-dominated cosmology \citep{Planck13} with $\Omega_{M}$=0.308, $\Omega_{\Lambda}$=0.692, and H$_{o}$=67.8 \kms~ Mpc$^{-1}$. All magnitudes are on the AB scale unless otherwise stated.

\section{Archival \textit{HST} imaging}\label{sec:archival-HST}
The sources within our sample have a rich set of multi-wavelength space and ground-based data sets. To assist in our analysis and interpretation of distinct regions within these quasar host galaxies, we utilize high angular resolution images from the \textit{Hubble Space Telescope}. We download fully-reduced data from the Barbara A. Mikulski Archive for Space Telescopes (MAST). Table \ref{tab:HST-archive} list the archival HST observations used in this study.

\begin{table*}
\centering
\caption{Archival HST imaging \label{tab:HST-archive}}

\begin{tabular}{ccccc}
\hline\hline
Object & 
Proposal ID & 
Instrument & 
Filter &
Exposure time 
\\
& 
& 
& 
& 
(s)\\
\hline
3C446 & 12975 & ACS-WFC & F814W & 2200\\
3C298 & 13023 & WFC3-UV & F606W & 1100\\
3C268.4 & 13023 & WFC3-UV & F606W & 1100\\
4C09.17 &5393 & WFPC2   & F555W & 2100\\ 
3C9     &13945& ACS-WFC & F814W & 2040\\
\hline
\end{tabular}
\end{table*}

We construct a model of the PSF using stars in the vicinity of the quasar within the FOV of each instrument. Images centered on each star are extracted in a box region of roughly 5\arcsec x 5\arcsec. We then subtract the local background for each star and median combine the stellar images into a normalized ``master" PSF. This PSF is then re-scaled to the quasar's peak flux and subtracted out at the spatial location of the quasar. In cases where the quasar was saturated, we scale the flux in the diffraction pattern of the PSF. 

\section{Black hole mass measurement}\label{sec:bh_masses}
Blackhole masses are calculated using the broad-\ha line width and luminosity using the scaling relation from \cite{Greene05} for a single epoch SMBH mass estimate. We describe the details of the nuclear spectrum fitting in \cite{vayner19a}, which comprises of multi Gaussian models with a broad component for the BLR emission, a narrow Gaussian for the narrow-line region, and an intermediate width Gaussian for the case where there is an outflow. We use the flux and width of the broadest Gaussian to compute the black hole mass. For 3C9, 3C298, there are strong telluric/filter transmission issues that prevent accurate measurement of the FWHM for the emission line. For these targets, we use the Mg II single epoch black hole mass estimate from \cite{Shen11}. The black hole masses are provided in Table \ref{tab:sample}. We assume an uncertainty of 0.4 dex on the SMBH masses.

\begin{deluxetable*}{lcclll@{\extracolsep{-10pt}}l}
\tiny
\tablecaption{QUART Sample properties \label{tab:sample}}
\tablehead{
\colhead{Name} & 
\colhead{RA} & 
\colhead{DEC} &
\colhead{z} &
\colhead{L$_{\rm bol}$} &
\colhead{L$_{\rm 178 MHz}$} &
\colhead{M$_{\rm BH}$} \\
\colhead{} & 
\colhead{J2000} &
\colhead{J2000} & 
\colhead{} & 
\colhead{($10^{46}$ \ergs)} & 
\colhead{($10^{44}$ \ergs)} & 
\colhead{\msun}}
\startdata
3C 9 & 00:20:25.22  & +15:40:54.77 & 2.0199 & 8.17$\pm$0.31 & 9.0 & 9.87 \\
4C 09.17 & 04:48:21.74  & +09:50:51.46 & 2.1170 & 2.88$\pm$0.14 &2.6 & 9.11 \\
3C 268.4 & 	12:09:13.61 & +43:39:20.92 & 1.3987 & 3.57$\pm$0.14 & 2.3&9.56 \\
7C 1354+2552 & 	13:57:06.54 & +25:37:24.49 & 2.0068 & 2.75$\pm$0.11 & 1.4& 9.86 \\
3C 298 & 14:19:08.18 & +06:28:34.76  & 1.439 & 7.80$\pm$0.30 & 12 &9.51 \\
3C 318 & 15:20:05.48  & +20:16:05.49 & 1.5723 & 0.79$\pm$0.04 & 4.0 &9.30 \\
4C 57.29 & 16:59:45.85 & +57:31:31.77 & 2.1759 & 2.1$\pm0.1$ & 1.9 & 9.10 \\
4C 22.44 & 17:09:55.01 & +22:36:55.66 & 1.5492 & 0.491$\pm$0.019 & 0.6 &9.64 \\
4C 05.84 & 22:25:14.70  & +05:27:09.06 & 2.320 & 20.3$\pm$1.00& 4.5 &9.75 \\
3C 446 & 22:25:47.26 & -04:57:01.39 & 1.4040 & 7.76 & 4.4 & 8.87 \\
4C 04.81 & 23:40:57.98 & +04:31:15.59 & 2.5883 & 0.62$\pm$0.02 & 9.3 & 9.58 \\
\enddata
\end{deluxetable*}

\section{Distinct regions within each quasar host galaxy}

In this section we outline how we define various regions within the data cube of each individual object.

\subsection{Spatially-Resolved Dynamically ``Quiescent" Regions}\label{sec:regions}
In the first survey paper, we outline our methodology for fitting the emission lines in individual spaxels of our data cubes. From these fits, we derive integrated intensity, velocity, and velocity dispersion maps. The errors on the radial velocity and dispersion maps come directly from the Least-squares Gaussian model fit. The flux map's errors come directly from integrating a noise spectrum in quadrature over the same wavelength range where the emission line is integrated. The fits are presented in the appendix of \citep{vayner19a}. Here we utilize the radial velocity and dispersion maps to select regions with low-velocity dispersion to search for gas in gravitational motion and search for regions where star formation may have recently happened.

We define a dynamically ``quiescent" region of our data set that contains gas with a velocity dispersion ($V_{\sigma}$) less than 250 \kms. A quiescent region that belongs to the host galaxy of the quasar must have a radial velocity $<400$ \kms as we expect the maximum rotational velocity for a given host galaxy to be at most 400 \kms. The maximum rotational velocity found for the most massive galaxies studied with IFS at z$\sim$2 is about 400 \kms \citep{Forster18}. We define gas with $V_{r}>|400|$ \kms and $V_{\sigma}<$ 250 \kms belonging to a merger system. A system is defined as a merger if there are components with $V_{r}>|400|$ \kms or more than one distinct kinematic component. For example, in the 3C298 system, two galactic disks are found to be offset by less than 400 \kms. All radial velocity and velocity dispersion measurements are relative to the redshift of the quasar. The redshifts for the individual quasars are calculated in \cite{vayner19a} and are taken from the fit to the narrow-line region. For sources with no spatially unresolved narrow emission, we use the redshift of the broad-line region. We label quiescent regions in the following manner: source name + direction + component A or B where A = component associated with the quasar, B = component associated with the galaxy merging with the quasar host galaxy. We follow these with a one or two-word comment about the region. Examples of description words are clump, diffuse, or tidal feature. Where clump referrers to a typical few kpc in size compact ionized emission typically seen in high redshift star-forming galaxies. Diffuse referrers to gas that has a surface density of less than typical clumpy star-forming regions. A tidal feature refers to ionized gas associated with a tidal tail in a merging system, containing both diffuse and clumpy ionized gas morphology.

For each dynamically quiescent region, we construct a 1D spectrum by integrating over its spaxels. We show an example of this for 4C09.17 in Figure \ref{fig:4C0917_all}, spectra of distinct regions for the rest of the sources are presented in the appendix (Figures 11-18). The emission lines in each spectrum are fit with multi-Gaussian profiles. In these plots, we also present the outflow regions from \citep{vayner19a}, to illustrate the location of dynamically quiescent regions relative to turbulent regions in these quasar hosts. From these fits, we derive integrated intensity and velocity dispersion that are presented in Tables \ref{tab:fluxes} and \ref{tab:SFR-Met}.

\begin{figure*}
    \centering
    \includegraphics[width=7in]{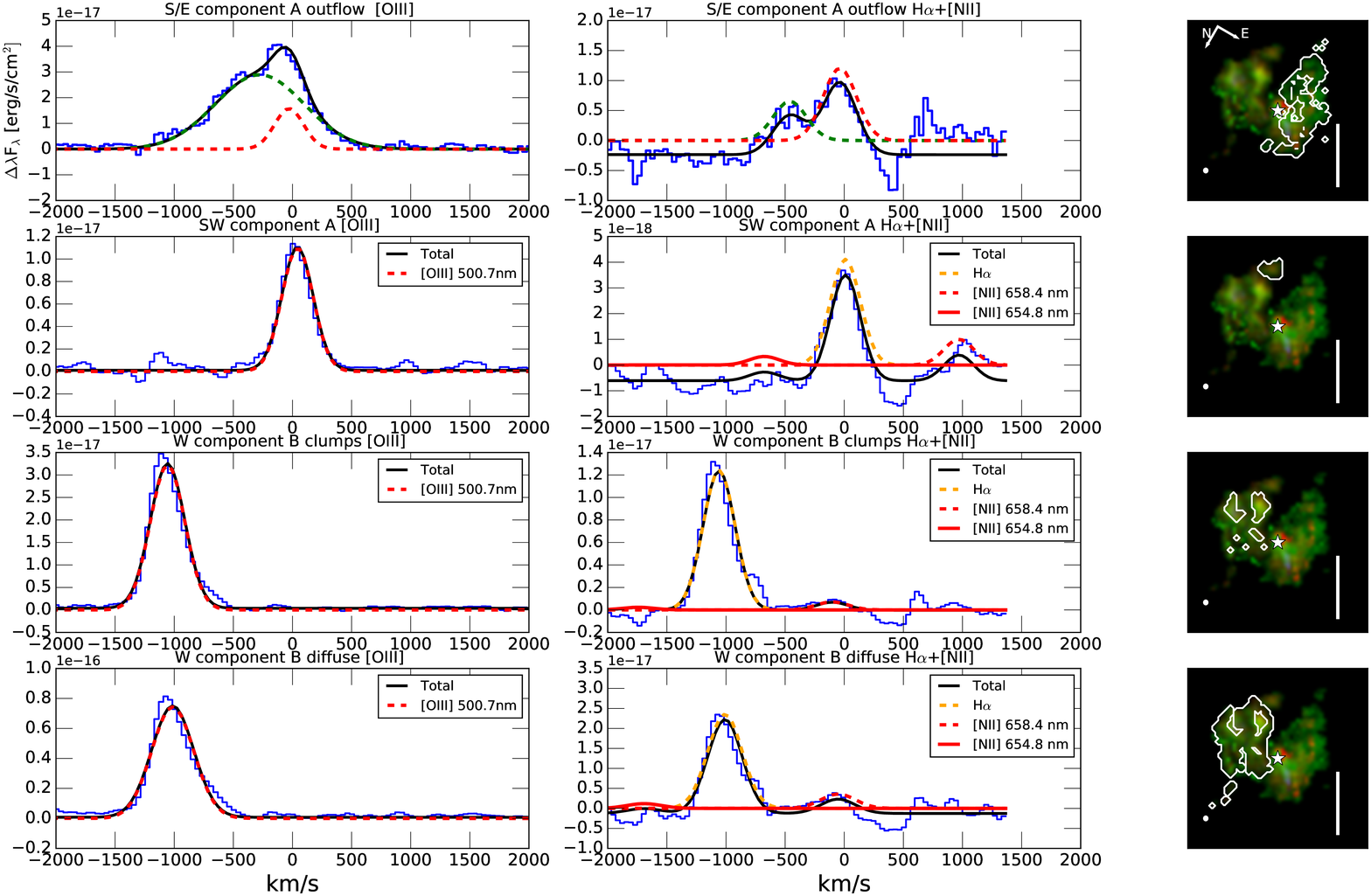}
    \caption{On the left, we present the spectra of distinct regions and fits to individual emission lines for the 4C09.17 system. On the right, we present the three-color composite where \ha is color-coded to red, \oiii to green and \nii to blue. The contour outlines the spatial location of the region. The bar on the right in each stamp represents 1\arcsec or approximately 8.6 kpc at the system's redshift.}
    \label{fig:4C0917_all}
\end{figure*}

\begin{deluxetable*}{llccc}
\tablecaption{Fluxes of distinct ``dynamically quiescent" regions in individual sources \label{tab:fluxes}}
\tablehead{\colhead{Source}&
\colhead{Region}&
\colhead{$\rm F_{[OIII]}$}&
\colhead{$\rm F_{H\alpha}$}&
\colhead{$\rm F_{[NII]}$}\\
\colhead{}&
\colhead{}&
\colhead{10$^{-17}$ \ferg}&
\colhead{10$^{-17}$ \ferg}&
\colhead{10$^{-17}$ \ferg}
}

\startdata
3C9     & SE-SW component A & 199$\pm$20 & 65$\pm$7 & 21$\pm$2 \\
        & N component B  & 127$\pm$13& 40$\pm$4 & 15$\pm$1\\
4C09.17 & SW component A           & 9.55$\pm$0.98 & 3.37$\pm$0.35 & 1.32$\pm$0.2\\
        & W component B clumps & 26$\pm$3 & 10$\pm$1 &0.77$\pm$0.13\\
        & W component B diffuse & 92$\pm$9&25$\pm$2&3.5$\pm$0.4\\
3C268.4 & SW component B           & 245$\pm$25& 51$\pm$5&9$\pm$1\\
7C 1354+2552 & component A      & 46$\pm$1  & 12$\pm$1&--\\ 
             & E component B       & 6.2$\pm$0.6  & 4.7$\pm$0.5&$<$0.7\\ 
3C298   & SE component B ENLR  & 649$\pm$65 & 188$\pm$20&65$\pm$7\\
        & SE component B tidal feature  & 55$\pm$5 &20$\pm$2 &3.6$\pm$0.5\\
4C57.29 & NE component A            &26$\pm$3 &-- &-- \\
        & N component A/B(?)        &12$\pm$1 &-- &--    \\
4C22.44 & N,S component A           &54$\pm$5 & 25$\pm$2 & 3.5$\pm$0.3\\
4C05.84 & SW component A clump     &7.7$\pm$0.8 &3.3$\pm$0.3&0.48$\pm$0.05\\
3C446   & NW component A tidal feature & 11$\pm$1 &5.9$\pm$0.6 & $<$0.15\\
        & E-W component B & 132$\pm$10 & 48$\pm$4 & 6.9$\pm$1.0 \\
\enddata
\end{deluxetable*}

\subsection{Spatially unresolved narrow-line regions}

We search for narrow spatially unresolved emission in each object. To do so, we first subtracted a model of the extended emission from our fits to each emission line in individual spaxels. We then perform aperture photometry on the spatially unresolved emission and extract a spectrum. The emission lines are fit with multiple Gaussian profiles. The fluxes of the narrow emission ($\sigma<250$ \kms) lines from unresolved regions are presented in Table \ref{tab:fluxes-unrsolved}. For sources where no unresolved narrow emission line is detected, we place a 1 sigma upper limit on the line flux. Based on the average angular resolution of about 0.1\arcsec, the unresolved narrow line emitting regions' sizes are $<$ 1 kpc.  

\begin{deluxetable*}{lccc}
\tablecaption{Fluxes of spatially unresovled narrow emission line regions in individual sources \label{tab:fluxes-unrsolved}}
\tablehead{\colhead{Source}&
\colhead{$\rm F_{[OIII]}$}&
\colhead{$\rm F_{H\alpha}$}&
\colhead{$\rm F_{[NII]}$}\\
\colhead{}&
\colhead{10$^{-17}$ \ferg}&
\colhead{10$^{-17}$ \ferg}&
\colhead{10$^{-17}$ \ferg}
}

\startdata
4C09.17 & 52$\pm$5 & 30$\pm$3 & 102$\pm$1\\
3C268.4 & 649$\pm$70& 239$\pm$20&76$\pm$8\\
4C22.44 &1521$\pm$200 & 102$\pm$10 & 3.5$\pm$0.3\\
3C318 & 35$\pm$4 & 66 $\pm$7 & 5$\pm$ 1 \\ 
3C446 & 11$\pm$1 &5.9$\pm$0.6 & $<$0.15\\
7C1354 & 65$\pm$10 & $<$4 & $<$4\\
4C57.29 & $<12$& $<$5 & $<$ 10 \\
4C04.81 & $<3$ & $<2$ & $<2$  \\
4C05.84 & $<0.9$&$<1$ &$<1$\\ 
\enddata
\end{deluxetable*}

\section{Nebular Emission Line Diagnostics and Sources of Gas Excitation}\label{sec:BPT}
In this section, we explore the photoionization mechanism in all distinct regions of each quasar host galaxy. The Baldwin, Phillips $\&$ Terlevich (BPT) diagram is used to differentiate between different gas photoionization sources \citep{Baldwin81}. Here, we use the \logohb and \loghn~line flux ratios to distinguish heating from young stars, AGN, and shocks.

To construct the BPT diagram for our sources, we integrated each emission line over the same velocity width ($\Delta$V) and velocity offset relative to the redshift derived from the \oiii emission line at each spaxel. We integrate the maps relative to \oiii since it is typically the brightest emission line in any given spaxel. The higher signal-to-noise \oiii emission line leads to a smaller spaxel-spaxel variation in the radial velocity and dispersion maps, creating a more consistent \logohb and the \loghn~ratio between neighboring spaxels. We find that for the entire sample, the standard deviation on the \logohb ratio decreases by 0.2 dex compared to when integrating the cubes relative to the \ha line. 

A resolved BPT diagram allows us to investigate the source of ionization throughout each quasar host galaxy. Due to sensitivity and, in some cases, wavelength coverage, we cannot create an integrated emission-line map for \hb~on a similar scale to \ha, \oiii, or \nii maps. For our BPT diagrams, we construct our \hb~map by assuming case B recombination (\hb=\ha/2.86) with a gas temperature of $10^4$ K and an electron density of $10^2$ cm$^{-3}$. Assuming other recombination cases and ISM conditions with reasonable temperatures and densities would not change our results by a significant amount as the ratios between \hb~and \ha would only change at most by a factor of $\sim$1.3 \citep{OsterbrocknFerland06}. For sources with the brightest extended emission and wavelength coverage of both \ha and \hb we find a maximum V band extinction of 1 mag, however in most cases, line ratios consistent with case B recombination. In regions where gas extinction is present, the \logohb ratios are preferentially lower.

\begin{figure*}
    \centering
    \includegraphics[width=6.5in]{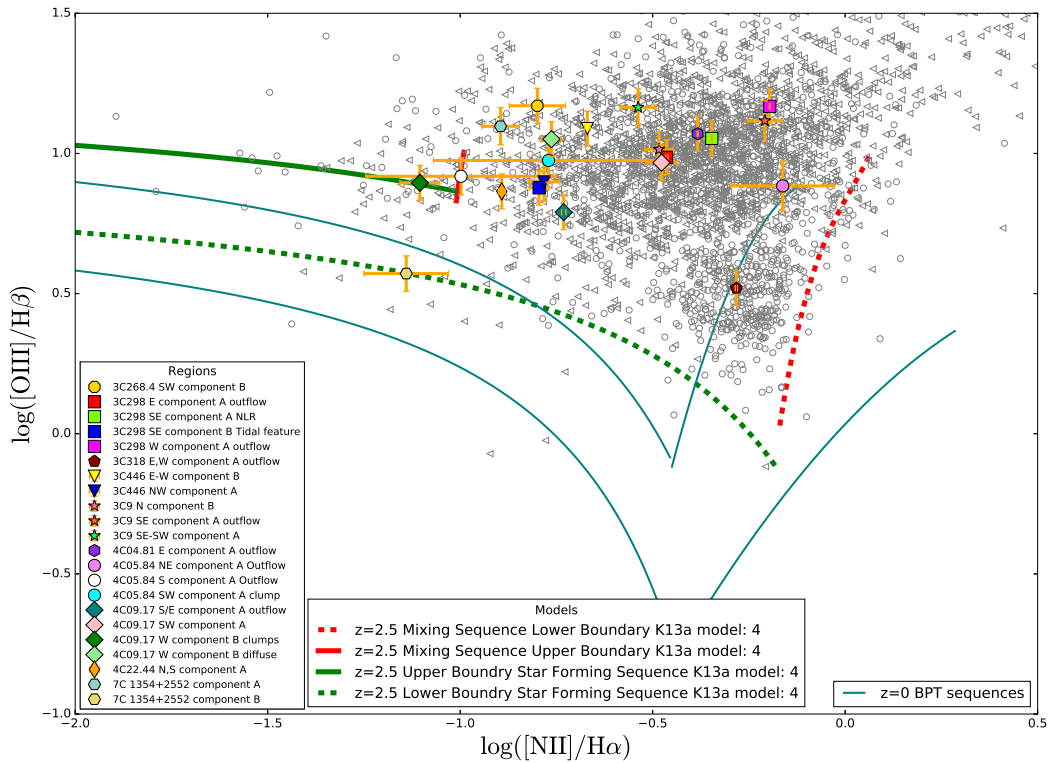}
    \caption{Line ratio diagnostics of individual resolved distinct regions. In grey, we plot the line ratios of individual spaxels where at least \oiii and \ha was detected at an SNR$>$3. Uncertainties on these line ratios are generally large; hence, we also integrate over all spaxels in individual regions to increase the SNR and lower the uncertainties on the line ratios. We show region-integrated line ratios with the colored symbols where each object has the same symbol, and each region has a different color. The names of the distinct region are present in the lower-left corner, and these match the names given in Table \ref{tab:fluxes}. We present the evolutionary models of the mixing and star-forming sequence with red and green curves from \cite{Kewley13a}. We show the upper limit of a sequence with a straight line and the lower boundary of each sequence with a dashed curve. Teal curves represent the bounds of the two sequences where the majority of the line ratio in low redshift galaxies fall. Our line ratios are consistent with a model where gas photoionized by the quasar is denser, has lower metallicity, and experiencing harder ionization compared to the gas photoionized by AGN in nearby galaxies.}
    \label{fig:BPT_total}
\end{figure*}

Only spaxels where at least \ha and \oiii were detected are analyzed and presented here. Typically \nii is detected in far fewer spaxels compared to \ha and \oiii. For spaxels where only \ha and \oiii are detected, we calculate a limit on \nii by integrating a sky spectrum over the same velocity width as \oiii at the expected wavelength location of \nii. In Figure \ref{fig:BPT_total} we plot the ratios from each spaxel. Diamonds are regions where \nii, \ha, and \oiii were detected, and triangles are regions where only \ha and \oiii were detected with a limit on the \nii flux. A total of 3160 spaxels are plotted corresponding to 21 distinct galactic regions.

For each distinct regions identified in section \ref{sec:regions} and from \citep{vayner19a} we over plot their line ratios and label them with a star. Individual spaxels typically have high uncertainties in their ratios but tend to cluster together on the BPT diagram. Integrating over distinct regions and re-calculating the ratios from a high SNR spectrum confirms that region's true line ratio.

\begin{figure*}
    \centering
    \includegraphics[width=6.5in]{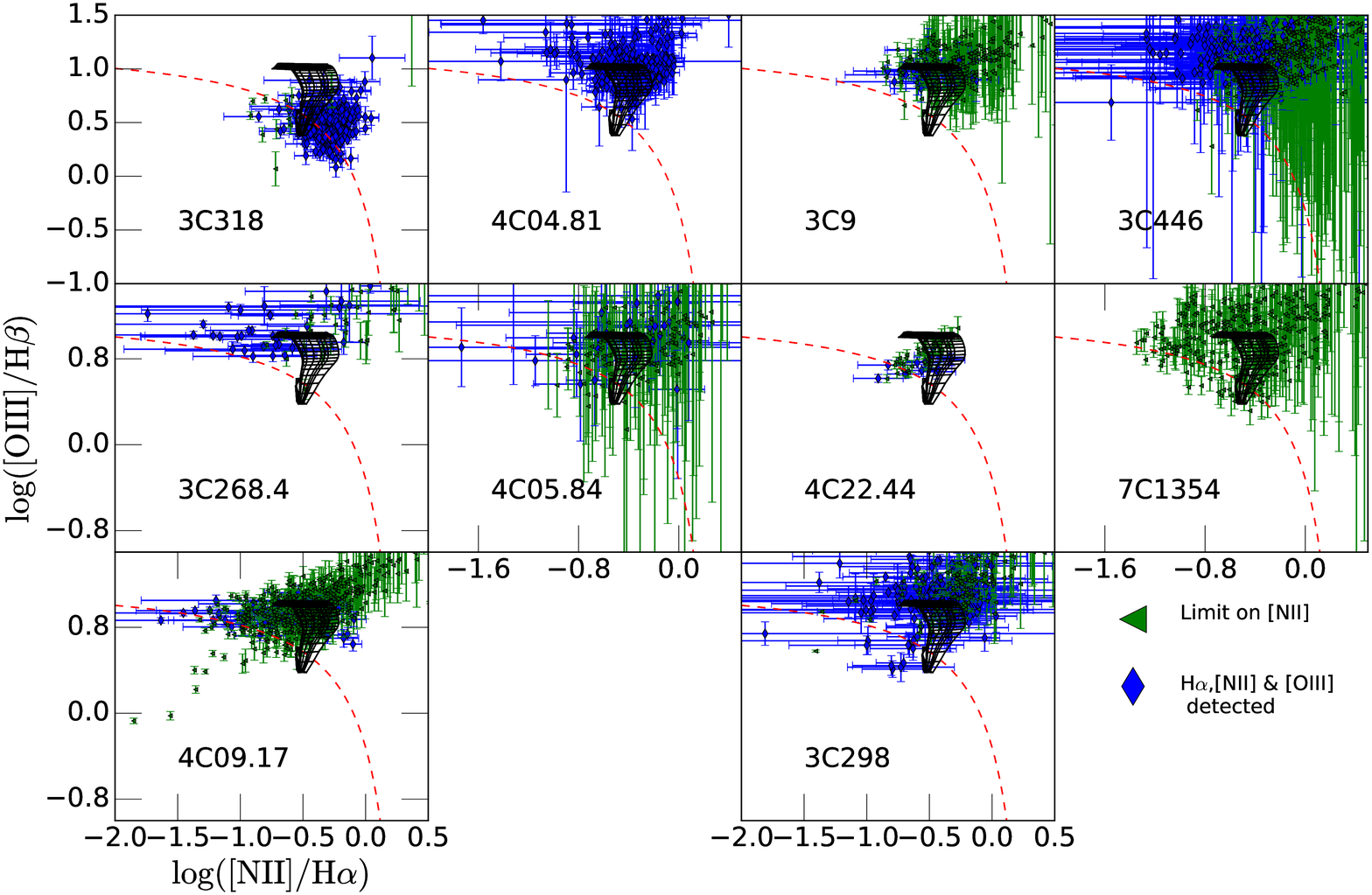}
    \caption{We present line ratio diagnostics for spaxels in each source where at least \oiii and \ha were detected at SNR great than 3. We show the uncertainties on the line ratio, which were omitted from figure \ref{fig:BPT_total} to conserve space. The dashed red line in each panel shows the theoretical separation between gas photoionized by star formation and AGN or shocks from \cite{Kewley13a}. Points above the line are photoionized by the quasar, while regions below are photoionized by O and B stars. Solid black mesh represents the location of radiative shocks from the astrophysical plasma modeling code MAPPINGS V \citep{Alarie19}. The shock model uses solar abundances from \cite{Gutkin16}. Either shocks or the quasar photoionizes the majority of the gas within our systems.}
    \label{fig:all_sources_BPT_diagram}
\end{figure*}

To conserve space, we do not over-plot the error bars on points from individual spaxels in Figure \ref{fig:BPT_total}, we only show the error bars of ratios computed for integrated values of the distinct regions. In Figure \ref{fig:all_sources_BPT_diagram}, we plot points of individual spaxels along with the error bars.

\subsection{Ionization Diagnostic Models}

We find that a large portion of our line ratios values lies outside the two typical sequences of the BPT diagram (Figure \ref{fig:BPT_total}). At a fixed \loghn~nearly, all values are above both the local mixing and star-forming sequence. At a fixed \logohb value, nearly all values are outside the local mixing sequence. A large portion of points falls between the star-forming and mixing sequence, with relatively high \logohb values. Metallicity, electron density, the hardness of the ionization field, and the ionization parameter determines the location of the galaxy/region on a given sequence. With changing conditions in the ISM between z=0 and the median redshift of our sample, the locus of both the star-forming and mixing sequence can change locations \citep{Kewley13a}. 

Galaxies at a fixed stellar mass have lower metallicities at high redshift compared to galaxies today \citep{Yuan13}. Near the peak of the star formation rate density at z$\sim1-3$, the ISM conditions and star formation histories of star-forming galaxies may differ from local systems. Star formation appears to happen in denser environments in the distant Universe with higher electron densities \citep{Sanders16}, akin to conditions seen in local ULIRGs. According to \cite{Steidel14, Strom17} ISM in high redshift galaxies experiences harder ionization at a fixed N/O and O/H abundances than z=0 star-forming galaxies. On the other hand, galaxies at higher redshift have elevated N/O rations \citep{Shapley15}. Taken together, \cite{Kewley13a} shows that such changes in ISM conditions can alter the location of the star formation sequence between z=0 and z=2.5. Notably, the combination of harder ionization, electron density, and ionization parameter can shift the locus of the star-forming sequence to higher \loghn~and \logohb values. It appears that UV/emission line selected galaxy samples tend to show a more significant shift from the SDSS star formation locus, as evident in a large sample of 377 star-forming galaxies explored by \cite{Strom17}. Nearly all their galaxies have a higher \logohb value at a fixed \loghn~compared to local galaxies studied in SDSS. Various galaxy selection techniques may lead to samples of galaxies with inherently different ionization properties. However, the overall conclusion from studying star-forming galaxies in the distant Universe is that the line ratios of these systems lie on different star formation locus compared to the local Universe.

Changes in the ISM conditions of distant galaxies may also lead to changes in the location of the mixing sequence. \cite{Kewley13a} and \cite{Groves06} show that for galaxies with lower metallicities, the mixing sequences shifts to lower \loghn~values with relatively small changes in the \logohb value. We explore the various evolutionary models of the star-forming and mixing sequence with redshift and ISM conditions proposed by \cite{Kewley13a}. The best fit model to our sample is the one where the ISM of high redshift galaxies have more extreme condition (higher electron density, harder ionization field, and larger ionization parameters) and the metallicity of the gas photoionized by the quasar is at a lower metallicity compared to the gas ionized by local AGN in the SDSS sample. The median \loghn~value is about 1.0 dex lower than that of the mixing sequence at z=0. If the primary source in the shift of the mixing sequence from z=0 to z=1.5-2.5 is a change in the gas phase metallicity, then the gas photoionized by the quasar in our sample has a metallicity a 1/2-1/5 of that in narrow line regions of z=0 AGN on the \cite{Kewley02} metallicity scale. One of the consequences of the shift in the mixing sequence is that it becomes harder to distinguish between gas photoionized by AGN vs. star formation, especially in systems with potentially multiple ionization sources. Changes in the photoionization condition likely also play a role in the offset from the local mixing sequence. In a sample of local type-1 quasars, \citep{Stern13} found that systems with higher bolometric luminosities and higher Eddington ratios are systematically offset to lower \loghn ratios.

Most of the gas in our quasar host galaxies lies in the mixing sequence where the gas is photoionized by a combination of quasar ionization and radiative shocks. In Figure \ref{fig:all_sources_BPT_diagram}, a significant fraction of the points in individual objects match the predicted location of radiative shocks on the BPT diagram. The radiative shock models assume solar metallicity and a preshock density of 215 \eden. With the present data, it is difficult to distinguish the percentage of photoionization from shocks vs. AGN. However, given the overlap of both line ratios and kinematics with shock models, we cannot rule them out to contribute to gas photoionization.

\begin{figure*}
    \centering
    \includegraphics[width=6.0in]{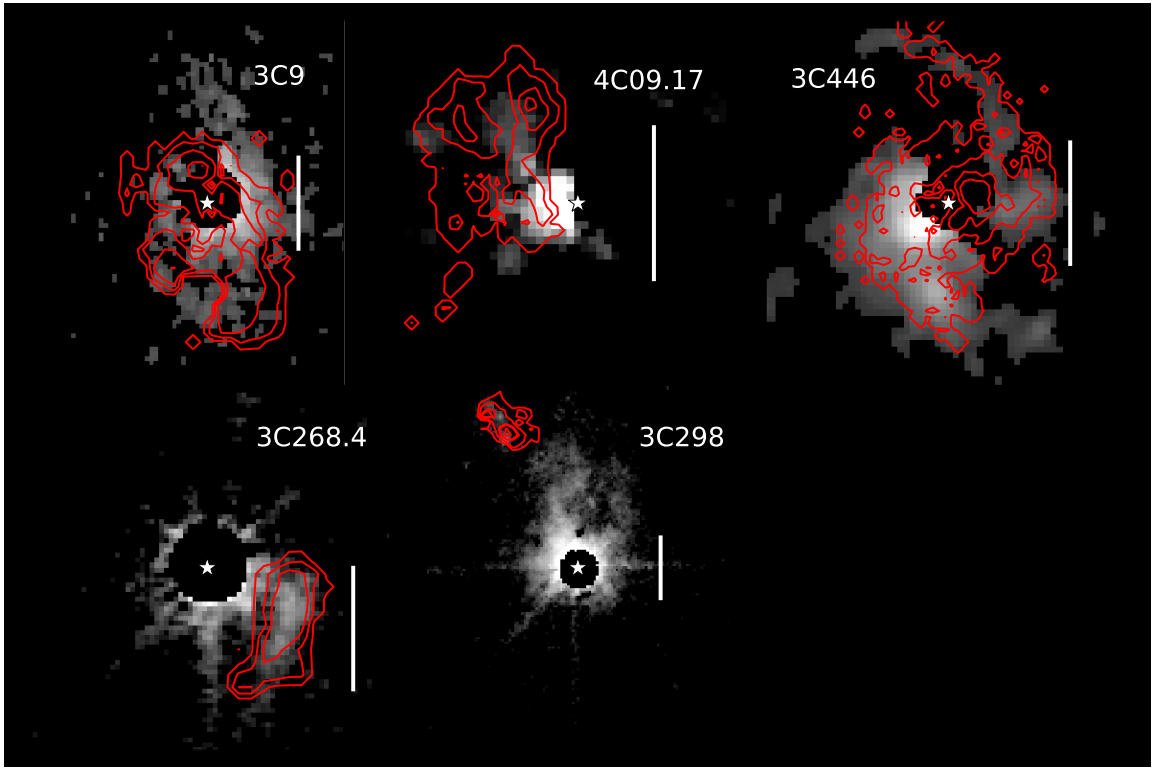}
    \caption{Detection of dynamically quiescent regions in archival \textit{Hubble Space Telescope} observation. In the background, we show PSF-subtracted images of rest-frame UV emission in the quasar host galaxy. Overlaid in contours is the extended \ha emission of the dynamically quiescent regions detected with OSIRIS. Note the similarities in both morphology and extent, indicating massive young stars photoionize at least a portion of the gas. The bar represents a spatial scale of 1\arcsec~or about 8.5 kpc.}
    \label{fig:quiescent-hst}
\end{figure*}

A number of our distinct regions appear to have low \loghn~values ($<$0.5) with low-velocity dispersion gas (V$_{\sigma}<250$ \kms). Morphologically these regions appear to be clumpy in their \ha maps reminiscent of typical star-forming regions in galaxies at $z>1$. The line ratios of these points do not coincide with regions of fast or slow shocks photoionization on the BPT diagram \citep{Allen08, Newman14}. Archival HST data of 3C9, 3C298, 4C09.17, 3C268.4, 3C446 all showcase that the dynamically ``quiescent" regions in these galaxies have clumpy morphology in rest-frame UV continuum data, similar to those of star-forming galaxies at these redshifts. In Figure \ref{fig:quiescent-hst}, we overlay the \ha emission from dynamically quiescent regions onto archival HST observations at rest-frame UV wavelength. Combining these clues suggests that the quasar does not entirely photoionize the gas in these regions. The elevated \logohb in these regions compared to local and distant star-forming regions may be from the mixing of photoionization from massive stars and the quasar. There is some evidence for this based on the morphology of the ionized gas and their respective \logohb ratios. For example, in 4C09.17, we see that more diffuse emission with low-velocity dispersion tends to have a higher \logohb value compared to clumpier regions where there is evidence for recent star formation activity and potentially more shielding from the quasar radiation field.

Using the empirical star formation rate \ha luminosity relationship from \cite{Kennicutt98}, we convert the \ha luminosity of the distinct extended quiescent regions to star formation rates. Most likely, the majority of these regions are photoionized by a combination of AGN and star formation, hence the derived star formation rates are upper limits. Regions ``3C298 SE component B Tidal feature" and ``4C09.17 W component B clumps" have line rations most consistent with photoionization by O/B stars, the star formation rates derived in these regions are closer to their actual value. To partially address the contribution from AGN to photoionization in dynamically quiescent regions, we also derive star formation rates only using spaxles that fall within the line ratio error inside the star-forming region of the BPT diagram based on the \cite{Kewley13b} maximum separation between star formation and AGN photoionization. Generally, we find lower (1/2 - 1/10) star formation rates when using the BPT diagram criteria. We also measure the metallicities of these regions using the \cite{Pettini04} empirical gas-phase metallicity - \loghn~relationship. Given that \loghn~is elevated in the presence of an AGN/quasar ionization field, the metallicities for the majority of the regions are also upper limits. We also calculate the metallicity using theoretically derived chemical abundance calibration for narrow-line regions of AGN \citep{SB98}. We present quantitative values of these regions in Table \ref{tab:SFR-Met}, where we show the \ha luminosity of each quiescent region, along with the star formation rate, metallicities, and velocity dispersion. Since nearly all of the unresolved narrow line regions are consistent with quasar photoionization, we do not include them in Table \ref{tab:SFR-Met} with the exception of 3C318. In this object, the line ratios are consistent with star formation, indicating a nuclear starburst on scales $<$1 kpc. For cases where we do not detect any unresolved narrow emission, we place an average upper limit on the star formation rate. We do so if there is an ongoing nuclear starburst with a star formation rate below the sensitivity of OSIRIS at our given exposure times. We find an average star formation rate limit of 9 \myr for the four objects (4C05.84, 4C04.81, 4C57.29, and 7C1354) where no strong narrow nuclear emission was detected.

\begin{deluxetable*}{rrlllllll}
\tablecaption{Star formation rates and metallicities of distinct dynamically quiescent regions \label{tab:SFR-Met}}
\tablehead{\colhead{Source}&
\colhead{Region}&
\colhead{L$\rm_{H\alpha}$}&
\colhead{SFR \tablenotemark{a}}&
\colhead{L$\rm_{H\alpha_{BPT}}$}&
\colhead{SFR $\rm_{BPT}$\tablenotemark{b}}&
\colhead{12+log(O/H)}&
\colhead{12+log(O/H)}&
\colhead{\vsigmagas}\\
\colhead{}&
\colhead{}&
\colhead{10$^{43}$\ergs}&
\colhead{\myr}&
\colhead{10$^{43}$\ergs}&
\colhead{\myr}&
\colhead{PP04}&
\colhead{SB98}&
\colhead{\kms}
}
\startdata
3C9     & SE-SW A           & 2$\pm$0.2 & 160$\pm$16 &0.20$\pm$0.02 &15$\pm$2& 8.6 &8.6 & 173.1$\pm$25.7\\
        & N B           & 1.2$\pm$0.1 & 99$\pm$10&0.06$\pm$0.006 &5$\pm$1 & 8.6 &8.5 & 200.5$\pm$5\\
4C09.17 & SW A       & 0.12$\pm$0.01 & 9$\pm$1  &0.04$\pm$0.01 &3$\pm$1&  8.6 &8.5& 126.6$\pm$3\\
        & W B clumps & 0.35$\pm$0.04 & 28$\pm$3 & \tablenotemark{a} &\tablenotemark{a}&  8.2 &8.4 & 136$\pm$4.7\\
        & W B diffuse & 0.86$\pm$0.09 & 68$\pm$7 &\tablenotemark{a} &\tablenotemark{a}& 8.4 &8.5 & 146.2$\pm$7\\
3C268.4 & SW B         &   0.64$\pm$0.06 & 51$\pm$5 &0.20$\pm$0.02 &18$\pm$2& 8.5 & 8.5 & 144.6$\pm$5\\
7C 1354+2552 & A    & 0.37$\pm$0.04 & 29$\pm$3 &0.4$\pm$0.04 &33$\pm$3&  $<$8.5 &8.4 & 182.16$\pm$38.2\\
3C298        & SE B TT\tablenotemark{d} &0.3$\pm$0.03 & 22$\pm$2 &\tablenotemark{a} &\tablenotemark{a}& 8.5 & 8.4 & 109.6$\pm$5.5\\
3C318        & Nuclear      &1.1$\pm$0.1&88$\pm$9&\tablenotemark{b}&\tablenotemark{a}&$<$8.3 &$<$8.5 & 179.8$\pm$7.4 \\
4C22.44      & N,S A        &0.40$\pm$0.04 &32$\pm$3 &0.2$\pm$0.02 &20$\pm$2 &8.4 &8.4 & 184.6$\pm$6.5\\
4C05.84      & SW A clump        &0.14$\pm$0.01 & 11$\pm$1&0.05$\pm$0.01 &5$\pm$0.5 & 8.5 &8.4 & 198.7$\pm$16\\
3C446        & NW A TT\tablenotemark{d} & 0.07$\pm$0.001 & 6 $\pm$1&0.06$\pm$0.01 &5$\pm$1 &$<$8.4 &$<$8.4 & 167.9$\pm$0.7\\
             & E-W B              & 0.48$\pm$0.05 & 38$\pm$4&0.15$\pm$0.02&12$\pm$2  & 8.5 &8.5 & 204.3 $\pm$ 15\\
\enddata
\tablenotetext{a}{Star formation rate derived using the \ha luminosity of the entire distinct quiescent region}
\tablenotetext{b}{Star formation rate derived using spaxels that fall within the star formation sequence on the BPT diagram.}
\tablenotetext{c}{Value indistinguishable from the integrated value over the entire dynamically quiescent region}
\tablenotetext{d}{Tidal tail}
\end{deluxetable*}

\section{SMBH-galaxy scaling relationships}\label{sec:galaxy_dyn_mass_sigma}
In this section, we place our galaxies on the velocity dispersion and galaxy mass vs. SMBH mass plots, comparing their locations to the local scaling relations (\msigma and \mstellar). We calculate the SMBH masses from the broad \ha luminosity and line width using the methodology presented in \cite{Greene05}. The SMBH masses span a range of $10^{8.87-9.87}$ \msun. The velocity dispersions are taken from dynamically quiescent regions, while the galaxy masses are calculated from the virial equation and from modeling the radial velocity of targets with rotating disks and extracting a dynamical mass.

\subsection{Host Galaxy Velocity Dispersion}

\begin{figure*}
    \centering
    \includegraphics[width=6.5in]{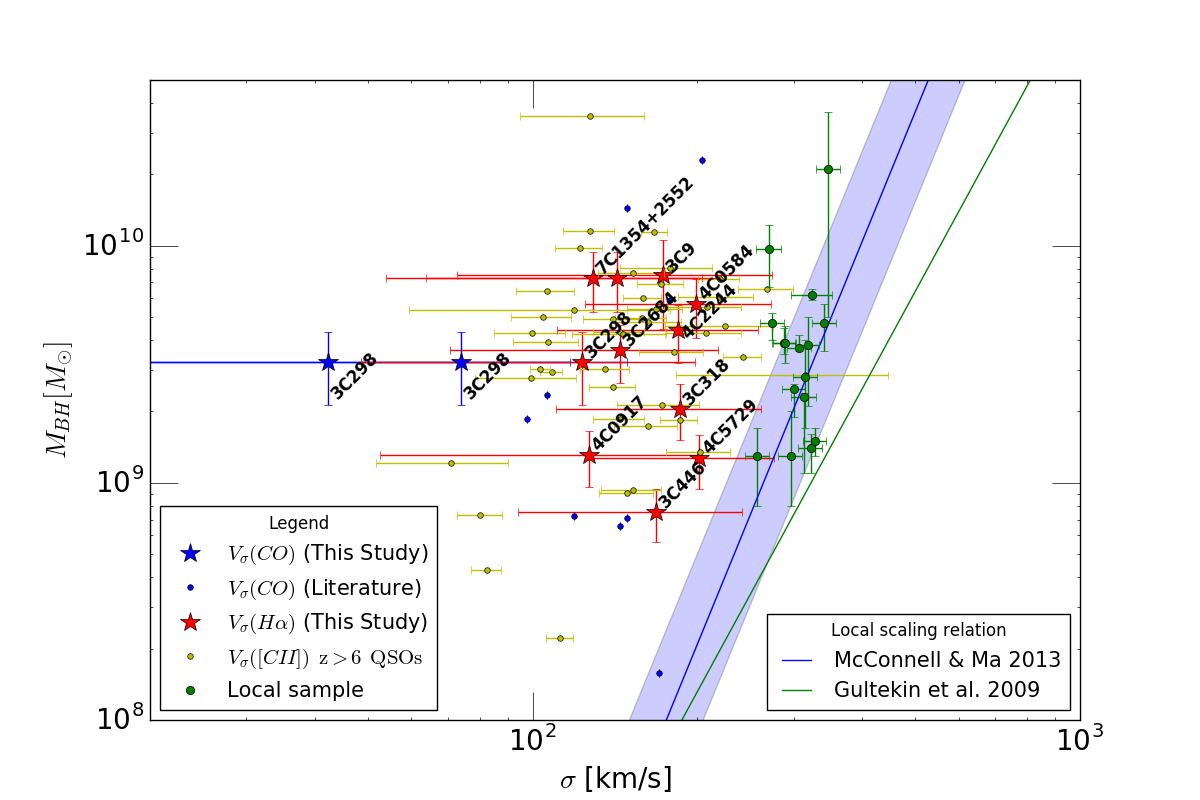}
    \caption{The location of our galaxies on the velocity dispersion vs. SMBH mass plot compared to the local \msigma relationship. We use the narrow \ha emission line velocity dispersion of dynamically quiescent regions as a proxy for the stellar velocity dispersion. Red stars are the measurements from our sample, where we measure the velocity dispersion from the narrow \ha emission line. We measure the black hole masses using the broad \ha line from the broad-line-region discussed in section \ref{sec:bh_masses}. The two blue stars represent the velocity dispersion measured in the disk of the host galaxy of 3C 298 and the tidal tail feature 21 kpc away from the quasar \citep{Vayner17}. Blue circles are quasars from the \cite{Shields06} sample, where they measure the velocity dispersion from CO emission lines. The yellow points are from quasars at z$>6$, where they measure the velocity dispersion from the 158 \micron~[CII] emission line \citep{Decarli18}.  Green points represent the local sample of all the bright cluster galaxies with SMBH greater than $10^9$ \msun taken from \cite{McConnell13}. The blue curve represents the best fit to the entire galaxy sample from \cite{McConnell13} with the blue shaded region represents the intrinsic scatter on the \msigma relationship from the fit while the green curve is the fit to the sample studied in \cite{Gultekin09}. We find a significant offset between the galaxies in our sample and local BCG and the general local \msigma relationship. This large offset indicates that the host galaxies appear to be under-massive for their SMBHs.}
    \label{fig:m-sigma}
\end{figure*}

We identify several dynamically quiescent regions within most of the quasar host galaxies in our sample. These regions show lower \loghn~line ratios and typically have clumpy morphology, reminiscent of the general star-forming regions seen in nebular emission and UV continuum in high redshift galaxies. In most galaxies, these regions lie away from any galactic-scale outflows. Hence their observed dynamics could be a probe of the galactic gravitational potential. These regions can be used to measure the velocity dispersion of our quasar host galaxies. In combination with the measured black hole masses, we can compare them to the local scaling relation between the SMBH mass and the velocity dispersion of the galaxy/bulge. In Figure \ref{fig:m-sigma}, we plot the mass of the SMBH presented in Table \ref{tab:sample} against the velocity dispersion of distinct quiescent regions measured with the \ha line. Also, we include the velocity dispersion measured from CO (3-2) emission for 3C 298 from \cite{Vayner17}. We find a significant offset from the local scaling relation between the SMBH mass and the velocity dispersion of the galaxy/bulge (\msigma) \citep{Gultekin09, McConnell13}. To address the issue that the velocity dispersion may be systematically lower in dynamically quiescent regions offset from the quasar (3C446) or regions where the surface area of the narrow emission is significantly lower than the outflow (4C09.17, 3C298), we recalculate the velocity dispersion in a larger aperture that includes outflows and narrow emission. We see no strong systematic difference in the velocity dispersion of the narrow gas. The source integrated narrow velocity dispersion for 3C298, 3C446 and 4C09.17 are 100.7 $\pm$ 16.6, 187.5 $\pm$ 1.0 and 144.0$\pm$10.0 \kms, respectively. In the nearby universe, the velocity dispersion is typically measured inside the effective radius of the galactic bulge. The difference within our observations is that we do not know the bulges' true sizes for our galaxies as we have no way to measure them with current data and telescope/instrument technology. However, the extents of the dynamically quiescent regions are in the range of the effective radii for bulges in massive ($10^{10.5-11.5}$ \msun) galaxies studied in the CANDELS survey \citep{Bruce14}.

\begin{figure*}
    \centering
    \includegraphics[width=6.5in]{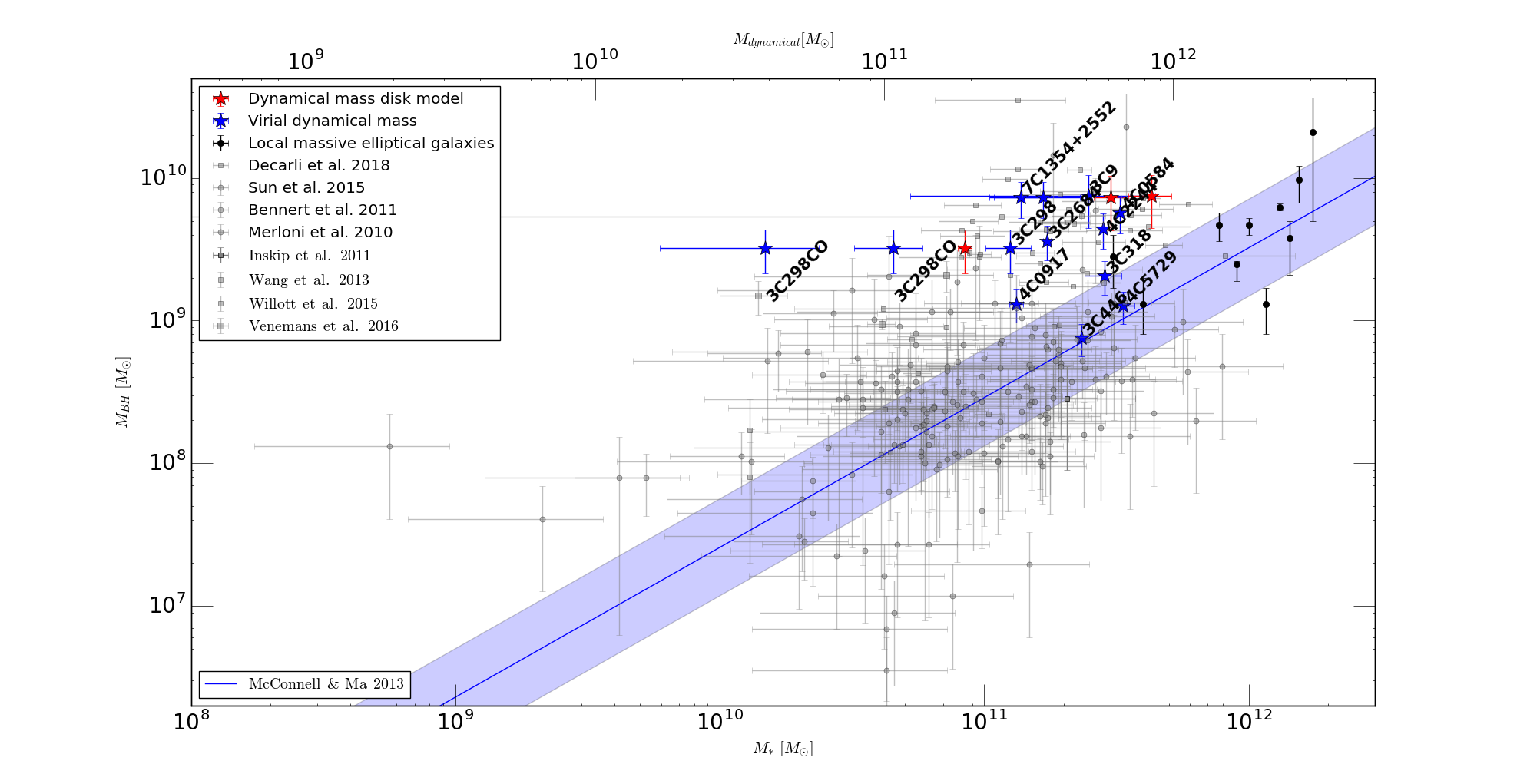}
    \caption{We present the location of individual galaxies compared to the local scaling relation between the mass of the SMBH and mass of the galaxy/bulge shown with a blue curve. Blue points represent systems with virial dynamical masses. Red points represent systems where we calculate the dynamical mass by modeling the radial velocity maps with an inclined disk model. Gray points show the location of galaxies at z$>0.5$, with lower SMBH masses and lower AGN luminosity compared to our sample. The blue curve represents the local scaling relationship as measured in \cite{McConnell13}, with the shaded region representing the intrinsic scatter. We find the majority of our points are offset from the local scaling relationship, outside the observed scatter.}
    \label{fig:m-stellar}
\end{figure*}

There have seen numerous discussions in the literature about whether the velocity dispersion measured from gas traces the stellar velocity dispersion. The gas and stars might not have the same uniform distribution, and winds can broaden the nebular emission lines. Furthermore, the line of sight absorption and emission lines from which the velocity dispersion is calculated are luminosity weighted subject to galactic dust extinction. Because of the different light distribution between stars and gas, the measured velocity dispersion can differ. These differences can lead to increased scattering in any correlation between \vsigmastellar and \vsigmagas. Data-sets that spatially resolve the gas and stellar components and have enough resolving power to separate multi-component emission from different regions (e.g., outflowing/in-flowing gas vs. galactic disk rotation) are important when making a comparison between \vsigmastellar and \vsigmagas. In \cite{Bennert18}, for a large sample of local AGN, they find when fitting a single Gaussian component to the \oiii emission line, they can overestimate the stellar velocity dispersion by about 50-100$\%$. Only by fitting multiple Gaussian components to account for both the narrow core and the broader wings of the \oiii line profile can they adequately match the velocity dispersion from the narrow component of the \oiii line to that of the stellar velocity dispersion. For their entire sample, the average ratio between the velocity dispersion of narrow Gaussian component and the stellar velocity dispersion is $\sim$1. The 1 $\sigma$ scatter on the ratio between $\sigma_{[OIII],narrow}$ and \vsigmastellar is about 0.32 with a maximum measured ratio of about a factor of 2 which translates to a scatter in $\Delta \sigma = \sigma_{[OIII]} - \sigma_{*}$ of 43.22 \kms with a maximum difference of about $\pm$100 \kms. However, only a few sources show such drastic velocity differences ($\sim2.5\%$ of the entire sample, 82$\%$ of the sources show $|\sigma_{[OIII]}-\sigma_{*}|<$ 50 \kms). When fitting for the \msigma relationship with the narrow \oiii emission as a proxy for stellar velocity dispersion, the resultant fit agrees with that of quiescent galaxies reverberation-mapped AGNs. These results indicate that for the sample as a whole \cite{Bennert18} finds that both the stars and gas follow the same gravitation potential. 

Given the \cite{Bennert18} results that demonstrates that the gas velocity dispersion can be used as a proxy for the stellar velocity dispersion, we follow a similar analysis using our IFS data sets to explore the location of our galaxies relative to the \msigma relation at high redshift. We attempted to the best of our ability to separate regions that contain galactic scales winds from those with more quiescent kinematics both spectrally and spatially with OSIRIS. Hence similar to \cite{Bennert18} we think that the measured velocity dispersions in quiescent regions are good tracers of the galactic potential on average. Throughout the paper we use the narrow velocity dispersion of \oiii and \ha emission lines of dynamically quiescent regions as a proxy for the stellar velocity dispersion. We still find a significant offset for our sample after applying the observed scatter in the difference between \vsigmastellar and \vsigmagas. This is also true when applied to the more distant quasar host galaxies studied with 158 \micron~[CII] emission. In nearby galaxies, there is a dependence on the velocity dispersion with the radius from the galaxy center \citep{Bennert18,Ene19}. However, based on the local galaxy observations, the velocity dispersion is unlikely to increase by $\sim$ 200 \kms that is necessary to bring the galaxies onto the local scaling relations.

Using N-body smoothed-particle hydrodynamics simulations \cite{Stickley14} examines how the stellar velocity dispersion evolves in a binary galaxy merger. At various stages in the merger (e.g., a close passage, nucleus coalescence), they measure the stellar velocity dispersion along $10^{3}$ random lines of sight. Near each close passage and during coalescence, they find that the scatter on the velocity dispersion significantly increases from  $\sim 5-11$ \kms to about 60 \kms with the average velocity dispersion a factor of $\sim$1.7 higher than after the galaxies have finished merging. For several sources in our sample (3C9, 3C298, and 3C446), the measured velocity dispersion might be higher than what it will be once the galactic merger is complete adding uncertainty due to projection effects. Following the simulations results, we add in quadrature an additional uncertainty on the velocity dispersion of 60 \kms given that the majority of our mergers are near coalescence or a close passage ($\Delta R < 10$ kpc).

It should be noted that this is near the maximum scatter seen in the simulations on $\sigma$. These simulations also find that for merging galaxies at their maximum separation, the measured velocity could be a factor of $\sim1.7$ times smaller compared to the final system. They find that for a 1:1 merger, the maximum separation after the first passage is 10-100 kpc, which is much larger than any separation that we find in our systems from observed projected separations and measured relative velocities. No obvious merging companions are found for 3C318, 4C22.44, or 4C05.84 hence for these systems, the mergers might be past their coalescence stage where the measured velocity dispersion is close to its final value, and the scatter due to the line of sight effects is minimal ($\sim$ a few \kms). However, we still apply an additional 60 \kms uncertainty in these regions.

Even after these corrections are made to approximate the stellar velocity dispersion from the \oiii emission lines in our sample, we still find that all of our systems are offset from the local scaling relation between the mass of the SMBH and the velocity dispersion of the bulge/galaxy. Given that we are dealing with relatively small sample size, we performed statistical tests to confirm the offset between the local scaling relation and our sample. We measure the offset between the observed and predicted velocity dispersion for the SMBH mass of our systems for each object. We use the local scaling relation fit from \citep{McConnell13}, and \ha measured SMBH masses. We construct a data set consisting of velocity differences. From bootstrap re-sampling of the velocity difference data set, we find that the average offset of 188.7 \kms is significant at the 3.25$\sigma$ level. Using Jackknife re-sampling similarly, we find that the offset is significant at the 3.3$\sigma$ level with the 95$\%$ confidence intervals of 154.4 \kms to  223.0 \kms on the velocity dispersion offset. Performing similar statistical tests on the \cite{Decarli18} sample, we find an average offset of 178.8 \kms with a significance of the shift at 2.7$\sigma$ and 2.8$\sigma$ for Jackknife and bootstrap re-sampling, respectively from the local relationship. We also measure the offsets of massive BCGs in the local Universe from the \msigma relationship. Using a two-sided Kolmogorov-Smirnov test, we can ask if the observed offsets of the local and high redshift data sets are drawn from the same continuous distribution. We find a p-value of 5.7$\times10^{-9}$, indicating that the two populations are not drawn from the same distribution. Applying the Kolmogorov-Smirnov test to the velocity dispersion offsets from our sample and in the higher redshift quasars, we find a p-value of 0.84, indicating that these two data sets could be drawn from the same continuous distribution. We find similar results by comparing the \cite{Shields06} sample at $z\sim2$ to our own and that of \cite{Decarli18}.

\section{Dynamical mass measurements}\label{sec:dyn-mass}
We can also test whether these systems lie off the local scaling relationship between the SMBH mass and the dynamical mass of the bulge/galaxy. First by using a virial estimator for the dynamical mass of the galaxy $\rm M_{virial} = \frac{C\sigma^{2}r}{G}$ where C=5 for a uniform rotating sphere \citep{Erb06b}. We assume 7 kpc for the radius, which is the median effective radius of massive quiescent galaxies in the local Universe \citep{Ene19}. Here $\sigma$ is derived from a Gaussian fit to the integrated spectra over the distinct region. For galaxies with multiple distinct regions, we derive two or more dynamical masses as there may be a dependence on the velocity dispersion as a function of position with the galaxy. For systems in a clear merger, the galactic component belonging to the quasar is used to estimate the dynamical mass since we are interested in the correlation between the SMBH and the velocity dispersion of the quasar host galaxy. 

For systems with velocity shear in the 2D radial velocity map, we fit a 2D inclined disk model to the kinematics data to measure the dynamical mass. The model is a 2D arctan function 

\begin{equation}
    V(r)=\frac{2}{\pi}V_{max}\arctan \Big( \frac{r}{r_{dyn}} \Big),
\end{equation}

\noindent where V(r) is rotation velocity at radius r from the dynamical center, $V_{max}$, is the asymptotic velocity, and $r_{dyn}$ is the radius at which the arc-tangent function transitions from increasing to flat velocity. The measured line-of-sight velocity from our observations relates to V(r) as

\begin{equation}
    V=V_{0} + \sin i\cos\theta V(r),
\end{equation}

\noindent where

\begin{equation}
    \cos\theta = \frac{(\sin\phi(x_{0}-x))+(\cos\phi(y_{0}-y))}{r}.
\end{equation}

\noindent Radial distance from the dynamical center to each spaxel is given by
\begin{equation}
    r = \sqrt{(x-x_{0})^{2}+\Big( \frac{y-y_{0}}{\cos i} \Big)^2},
\end{equation}

\noindent where $x_{0},y_{0}$ is spaxel location of the dynamical center, we quote the value relative to the centroid of the quasar, $V_{0}$ is velocity offset at the dynamical center relative to the redshift of the quasar, $\phi$ is position angle in spaxel space, and $i$ is the inclination of the disk. $V_{max}$ is not the true ``plateau" velocity of the galaxy's disk. $V_{max}$ can have arbitrarily large numbers, especially when $r_{dyn}$ is very small \citep{Courteau97}. To fit the data we use the MCMC code \textit{emcee}. We construct the model in a grid with a smaller plate scale than the observed data, which gets convolved with a 2D Gaussian PSF with an FWHM measured from the quasar PSF image. The image is then re-sized to the plate scale of the data. We construct the priors on each of the seven free parameters. The prior on $V_{max}$ is $300<V_{max}<1000$ \kms the prior on both $x_{0},y_{0}$ is the boundary of the FOV of the imaged area, the prior on the position angle is $0<\phi<2\pi$, the prior on the inclination angle is $0<i<\pi/2$, the prior on the radius is $0.5<r_{dyn}<10$ pixels and the prior on $V_{0}$ is $-100 < V_{0} <100$ \kms. We then sample this distribution with \textit{emcee}. We initialize 1000 walkers for each free parameter using the best fit values from \textit{leastsquares} fitting as the starting point, with a small random perturbation in each walker. We run MCMC for 500 steps starting from the perturbed initial value. The best-fit parameters, along with their confidence intervals, are presented in \ref{tab:disk-fit} for the quasar host galaxies of 7C 1354+2552, 3C9. For 3C 298 we do not see the disk in the ionized emission with the OSIRIS data, it is solely detected in CO (3-2) observations from ALMA, here we present the best fit values from \cite{Vayner17}. Also, we present $\Delta v_{obs}/2$, the average between the maximum and the minimum velocity along the kinematic major axis as determined by the position angle ($\phi$). We also present the intrinsic velocity dispersion ($\sigma_{0}$), measured along the kinematic major axis, towards the outskirts, away from the steep velocity gradient near the center of the disk.

\begin{table}
\centering
\caption{Best fit values for each inclined disk model parameter \label{tab:disk-fit}}
\begin{tabular}{llll}
\hline\hline
Parameters & 
7C 1354+2552 & 
3C9 &
3C298\\
\hline
$V_{max}$ [\kms] & 449.67$^{+0.24}_{-0.64}$ & 442.0$^{+23.9}_{-5.7}$ & 392$^{+65}_{-65}$ \\
$x_{0}$ [kpc] & -2.37$^{+0.04}_{-0.03}$&  0.5$^{+2}_{-1}$ &  0.43$^{+0.1}_{-0.1}$    \\
$y_{0}$ [kpc] & -0.93$^{+0.08}_{-0.08}$ & -4.8$^{+1.22}_{-1.5}$ &  0 $^{+0.1}_{-0.1}$    \\
$\phi$ [$^{\circ}$] & 75.68$^{+0.47}_{-0.48}$&74.10$^{+3.5}_{-35.4}$ &  5.3$^{+1.28}_{-1.28}$     \\
$i$ [$^{\circ}$] & 47.6$^{+0.8}_{-0.8}$& 47.1$^{+5.0}_{-3.7}$&  54.37$^{+6.4}_{-6.4}$     \\
$r$ [kpc] & $<$0.017 & 0.26$^{+0.49}_{-0.14}$ &  2.1$^{+0.9}_{-0.9}$     \\
$V_{0}$ [\kms] & -93.9$^{+1.2}_{-1.7}$&-9.22$^{+30.45}_{-86.46}$ & -13.0$^{+3.15}_{-3.15}$      \\
$\Delta v_{obs}/2$ [\kms] &309.84$\pm$20.47 & 370.84$\pm$45.4 & 150.0 $\pm$23.7\\
$\sigma_{0}$ [\kms] & 61.3$\pm$7.9 & 186.9$\pm$32.7 & 42.35 $\pm$ 12.68\\
\hline
\end{tabular}

\end{table}
In addition, we measure $V_{rot}/\sigma_{0}$ to gauge whether these systems are dynamically supported by rotation or dispersion. We measure a value of 6.8$\pm$1, 2.7$\pm$0.6, and 4.4$\pm$1.5 for 7C 1354, 3C9, and 3C298, respectively. In all systems, rotation dominates over the velocity dispersion for the dynamical support according to the criteria outlined by \cite{Forster18}, and henceforth the systems can be classified as true disks.

Assuming a spherically symmetric system, we can compute the total enclosed mass using the following formula: 

\begin{equation}
    M(R) = 2.33\times10^{5}rV_{r}^{2}/\sin(i)^{2}
\end{equation}

Where $V_{r}$ is the radial velocity, $i$ is the inclination angle from the disk fit. For the radial velocity we use $\Delta v_{obs}/2$. Similarly, we assume a radius that is the median value of nearby BCGs (7.1 kpc). The selected radius should give us an absolute upper limit on the dynamical mass of the galaxy/bulge as this radius is much larger than the typical size of a galactic bulge at this redshift and is larger than the observed extent of the galactic disks. The reason for choosing a larger radius is to address the case where the quasar host galaxy extends to a larger radius and is not captured in our OSIRIS observations because they are not sensitive enough to low surface brightness emission at larger separation from the quasar. Virial and dynamical masses are presented in \ref{tab:dynamical_masses}. However, it is not guaranteed that the extent of the ionized gas will match the stellar. We attempted to measure the size of the stellar continuum from the HST observations but were unsuccessful. Using the Galfit package, we were unable to constrain the radius due to the sources' complex morphologies and the increased inner working angle due to the quasars' brightness and saturated counts in the HST observations.

Due to the limited sensitivity of OSIRIS to lower surface brightness emission, we are missing an accurate measurement of the plateau velocity for the galactic disks at large separations from the quasar. Hence, our fitting routine is unable to constrain $V_{max}$ for 3C9 and 7C 1354. Also, it appears that the turn over radius is very small for these two systems, smaller than the resolution element of our observations. For this reason, we are unable to constrain the turn over radius, and we only provide a limit. For 7C 1354, there is a degeneracy between the maximum velocity, turn over radius, and inclination; thus, the values that we provide are those that give the smallest velocity residual.

\begin{figure*}
    \centering
    \includegraphics[width=6.5in]{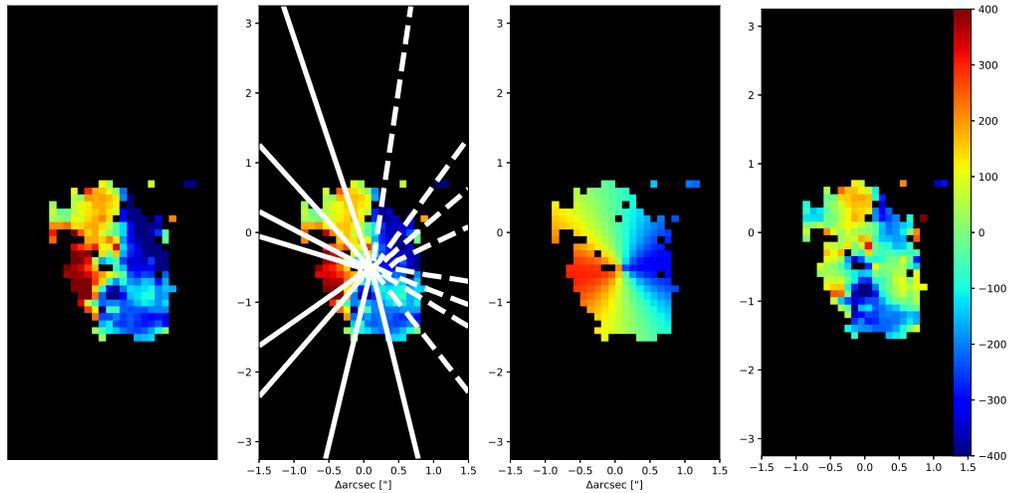}
    \caption{Fitting an inclined disk model to the radial velocity map of the 3C9 quasar host galaxy. Far-left we plot the isolated radial velocity structure belonging to the quasar host galaxy of 3C9, middle left shows the best fit model overlaid as contours on top of the radial velocity map, middle right is the best fit model. On the right, we plot the residuals. Larger blue-shifted residuals at $-1$\arcsec~south from the quasar are from the outflow (3C9 SE component A outflow A).}
    \label{fig:3C9_disk_fit}
\end{figure*}

\begin{figure*}
    \centering
    \includegraphics[width=6.5in]{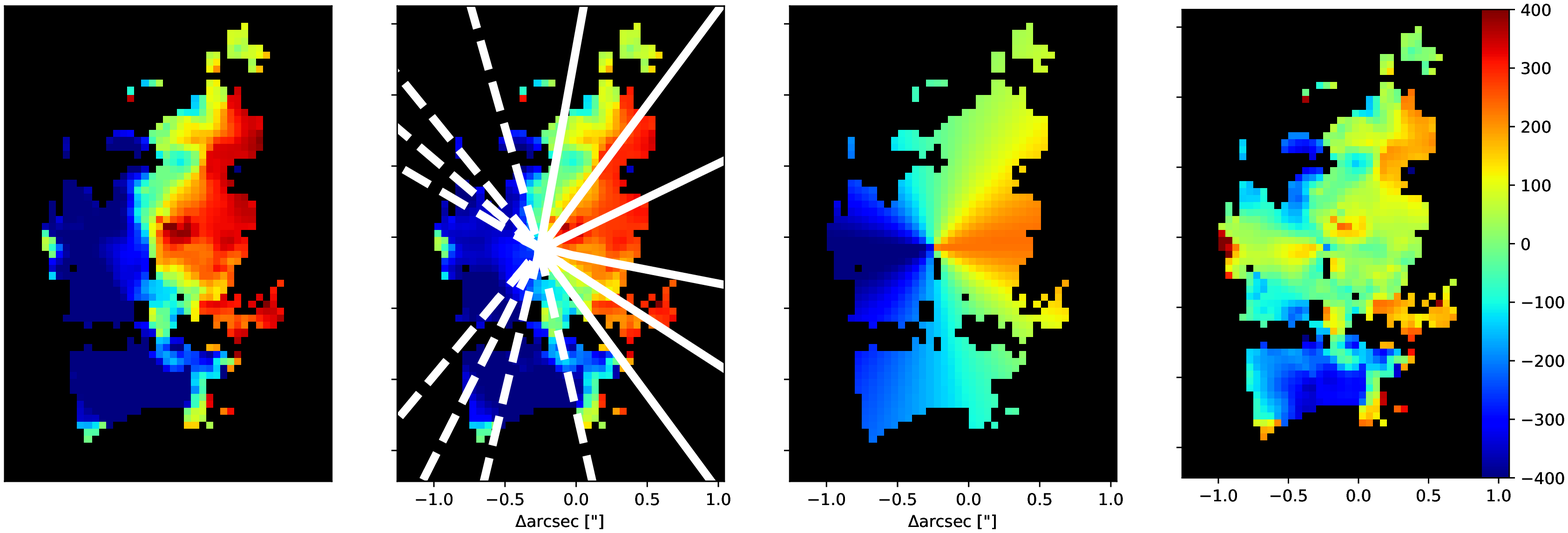}
    \caption{Fitting an inclined disk model to the radial velocity map of the 7C1354 quasar host galaxy. Far-left we plot the isolated radial velocity structure belonging to the quasar host galaxy of 7C1354, middle left shows the best fit model overlaid as contours on top of the radial velocity map, middle right is the best fit model, and on the right, we plot the residuals.}
    \label{fig:7C1354_disk_fit}
\end{figure*}

\begin{figure*}
    \centering
    \includegraphics[width=6.5in]{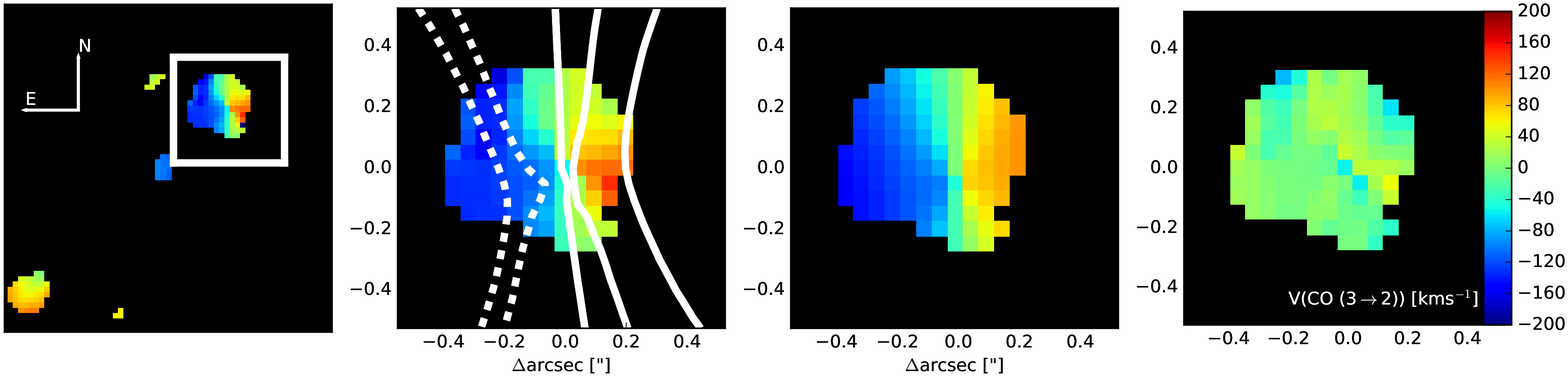}
    \caption{Fitting an inclined disk model to the radial velocity map of the 3C298 quasar host galaxy. Far-left we plot the isolated radial velocity structure belonging to the quasar host galaxy of 3C298, middle left shows the best fit model overlaid as contours on top of the radial velocity map, middle right is the best fit model, and on the right, we plot the residuals.}
    \label{fig:3C298_disk_fit}
\end{figure*}

\begin{table}[]
    \centering
    \caption{Virial and dynamical mass values.}
    \begin{tabular}{ccc}
    \hline
    \hline
    Source & Virial Mass & Disk-fit Dynamical mass\\
                     &$\times10^{11}$ \msun &$\times 10^{11}$ \msun\\
    \hline
    3C9              &2.5$\pm$0.7   &4.3$\pm$0.8\\
    4C09.17          &1.3$\pm$0.1   &--\\
    3C268.4          &1.4$\pm$0.1   &--\\
    7C1354+2552      &1.5$\pm$0.3   &3.0$\pm$0.4\\
    3C298            &0.45$\pm$0.13\tablenotemark{a} &0.6$\pm$0.1\\
    3C318            &2.9$\pm$0.5   &--\\
    4C57.29          &3.3$\pm$0.3   &--\\
    4C22.44          &2.8$\pm$0.1   &--\\
    4C05.84          &3.3$\pm$0.1   &--\\
    3C446            &2.3$\pm$0.1   &--\\
    
    \hline
    \end{tabular}
    
    \label{tab:dynamical_masses}
    \tablenotetext{a}{computed from CO 3-2 velocity dispersion}
\end{table}

Using the measured virial and disk fit dynamical masses and the SMBH masses, we can now compare our galaxies to the local \mstellar relationship. Not only are these galaxies offset from the local \msigma relationship, but we also find that these galaxies are on an average offset from the local \mstellar relationship. The galaxies need about an order of magnitude of stellar growth if they are to evolve into the present-day massive elliptical galaxies. 

We note that we have used two different methods for exploring the scaling relationship for galaxy mass vs. SMBH. Both the gas velocity dispersion method and dynamical measurement imply that the SMBH is over-massive compared to their host galaxies when exploring the local scaling relationship. It will be important to further compare these methods with larger samples, as well as future observations with the James Webb Space Telescope that will be able to directly measure the stellar velocity dispersion.

\section{Discussion}\label{sec:chapter4discussion}
Our survey aimed to study host galaxies of redshift 1.4 - 2.6 radio-loud quasars through rest frame nebular emission lines redshifted into the near-infrared.

We place distinct regions of each quasar host galaxy on the traditional BPT diagram (\logohb vs. \loghn). The majority of the points for our sources lie outside the two local sequences (the mixing and star-forming sequence). In section \ref{sec:BPT}, we introduce evolutionary BPT models from \cite{Kewley13a} that indicate changes in the photoionization and metallicity conditions of the gas can shift both of the star-forming and mixing sequences. We fit these models to our data and find that the best-fitting model is the one where the gas in our quasar host galaxies is at least two to five times less metal-rich compared to the narrow line regions of nearby (z$<$0.2) AGN. The best fit model also indicates that the gas is ten times denser compared to nearby galaxies. In Figure \ref{fig:BPT_total}, we show all of our points on the BPT diagram along with the best fit model. \cite{Kewley13b} studied a sample of star-forming galaxies and galaxies with AGN in the redshift range of 0.8$<$z$<$2.5. They also find that galaxies at z$>$2 show elevated line ratios on average outside the local star formation and mixing sequences. They find that normal ISM conditions similar to the SDSS sample transition to the more extreme conditions with elevated line ratios somewhere between redshift z=1.5 and z=2. This is an agreement with our results as the majority of our targets are at $z>1.5$.

High redshift radio galaxies also appear to show ISM conditions with metallicities that are lower compared to local AGN. In a study of a large sample of distant radio galaxies, \cite{Nesvadba17} finds that their gas-phase metallicities are at least half of that seen in local AGN. \cite{Nesvadba17} finds the same best-fitting model from \cite{Kewley13a} as we do for our sample to explain their observed nebular line ratios. The average \loghn~value of our sample seems to be lower than that of \cite{Nesvadba17}; this could be due to the lower metallicity of our sample. On the other hand, a different approach to how we compute our line ratios can cause the discrepancy. \cite{Nesvadba17} only presents source integrated line ratios, while we explore ratios of distinct regions because we typically have a factor of 5-10 better angular resolution due to adaptive optics and hence can resolve the different ionized/kinematics structures of our galaxies. In the majority of our sources, we see significant variations in \loghn~and \logohb values across each system, hence why we explore distinct regions. Line ratios from integrated spectra that include regions with various ionization sources and from multiple components of a merger system may shift towards higher \loghn, and \logohb values as the regions photoionized by the quasar/AGN tend to be brighter. Line ratios of galaxies with lower luminosity AGN compared to quasars/radio galaxies studied in \cite{Strom17} are nearly all outside the local mixing sequence. These points overlap with the location of our line ratios and that of the radio galaxy sample. The MOSDEF survey finds similar results for their AGN sample at a range of bolometric luminosities \citep{Coil15}. The ubiquity of elevated line ratios in host galaxies of AGN, meaning they are typically above the local mixing or star-forming sequence on the traditional BPT diagram (\logohb vs. \loghn), indicates that regardless of the active galaxy population selected at z$\sim$2 the conditions of the gas that is photoionized by an AGN are different from those in the local Universe.

Overall, this suggests that the ISM conditions in high redshift galaxies with AGN at a range of bolometric luminosities are different from those in local systems. The ISM conditions appear to be far more extreme with gas-phase metallicity lower than that of local AGN, suggesting an evolution in the ISM gas that is photoionized by AGN from z=0 to z=2.5. 

\subsection{Star formation and dynamically ``quiescent" regions in the host galaxies}

In 9/11 quasar host galaxies within our sample, we see the morphology of clumpy star-forming regions seen in other galaxies at these redshifts. These regions also typically show lower velocity dispersion and lower \loghn~values. We described them in more detail in section \ref{sec:regions}. These regions lie 1 - 21 kpc from the quasar and generally do not coincide with the location of galactic outflows. For sources with available HST imaging of rest-frame UV continuum, these regions appear bright and clumpy (see Figure \ref{fig:quiescent-hst}). Taking these two results together indicates that O and B stars could photoionize a non-negligible fraction of the gas in these clumpy regions. In section \ref{sec:BPT}, we derive an upper limit on their star formation rates and gas-phase metallicities. 

Taking this together, there is evidence for very recent star formation activity in 9/11 quasars within our sample. We find an average star formation rate of 50 \myr for the star-forming regions within our sample. The average dynamical mass of our quasar host galaxies of $\sim10^{11}$ \msun, indicates that the galaxies sit near the galaxy star formation rate - stellar-mass sequence at z$\sim$ 2 \citep{Tomczak16}. Using the average metallicity of 8.5 measured in dynamically quiescent regions and the average stellar mass of our sample indicates that our galaxies sit on the mass-metallicity relationship at z$\sim$2 \citep{Sanders15}.

Quasars at z$\sim$ 2 are found to reside in galaxies with a broad range of star formation rates, spanning from quiescent to star-bursting galaxies. However, our sample preferentially contains quasar host galaxies in a star-burst phase. High specific accretion rate AGN are more likely to be found in star-bursting galaxies with rates on or above the star formation rate - stellar-mass sequence in the distant Universe \citep{Aird19}. We selected to observe compact steep spectrum radio-loud quasars, this class of objects tend to contain younger AGN. One of the mechanisms to trigger a luminous AGN is through a massive gas-rich galaxy merger \citep{Treister12}. During the ongoing merger, the loss of angular momentum feeds gas into the centers of galaxies, providing fuel for both star formation and SMBH growth. Since we selected AGN that may have recently triggered, they are more likely to be in an ongoing merger, where star formation activity is enhanced. Indeed, about 7/11 of the quasar host galaxies in our sample are mergers. This can explain why our sample preferentially contains galaxies with active or recent star formation and rapid accretion onto the SMBH.

The measured star formation rates within our sample are significantly lower than those measured through dust emission in the far-infrared by the \textit{Herschel Space Observatory} \citep{Podigachoski15, Barthel17} for 4C04.81, 4C09.17, 3C318, and 3C298. The most likely explanation is that the quasar itself could partially heat the dust, \ha misses a significant fraction of the obscured star formation, or the dust traces a different star formation history. Interestingly for 3C298 and 3C318, where both high spatial resolution imaging of the dust and \ha emission is present, there is a significant misalignment between the maps. In places where we see evidence for recent star formation based on nebular emission-line ratios in 3C 298 and 3C 318, \cite{Barthel18, Barthel19} does not see any dust emission. For the case of 3C 298 in the location where we see recent star formation traced by \ha, we also detect a molecular reservoir; however, no dust emission is present there. Furthermore, in the places where dust emission exists in the case of 3C 298, the molecular gas at that location is stable against gravitational collapse and has been on a time scale longer than the propagation of the observed outflow. For the case of 4C09.17 and 4C04.81, no high-resolution dust maps are available. The dust emission could originate at any location within the $\sim$ 17\arcsec~Herschel SPIRE beam, which translates to a physical scale of about 150 kpc. Future high spatial resolution dust and molecular gas emission maps are necessary for proper comparison between the obscured and unobscured star formation traces and the molecular gas dynamics.

\subsection{Offset from local scaling relations}

The majority of our systems appear to be offset from both local scaling relationships between the mass of the SMBH and mass and the velocity dispersion of the bulge (see Figures \ref{fig:m-sigma}, \ref{fig:m-stellar}). To explain the large offset from the local \msigma and \mstellar relationship, we could invoke a significant error in the estimated SMBH masses. The bolometric luminosities of some of our quasars are far greater than those used for reverberation mapping in the nearby Universe, which is used in calibrating the single epoch SMBH mass \citep{Greene05}. The SMBH masses would have to be off by 2-3 orders of magnitude to explain the observed offsets. By assuming that the SMBH grows primarily through gas accretion, we can use the Eddington luminosity formula to estimate the SMBH mass. Given that our quasars are most likely not all accreting at or close to the Eddington limit, this derived mass is effectively a lower limit. 

\begin{equation}
  \rm M_{SMBH,min}  = \frac{L_{Eddington}}{1.26\times10^{38}} M_{\odot}
\end{equation}

\noindent For the derived bolometric luminosities in Table \ref{tab:sample} we find a range of minimum SMBH of 10$^{7.5-9}$\msun, consistent with what we measure from single epoch SMBH masses using the \ha emission line. Hence, ther is likely no significant error in our measured black hole masses.

\begin{figure*}
    \centering
    \includegraphics[width=6.5in]{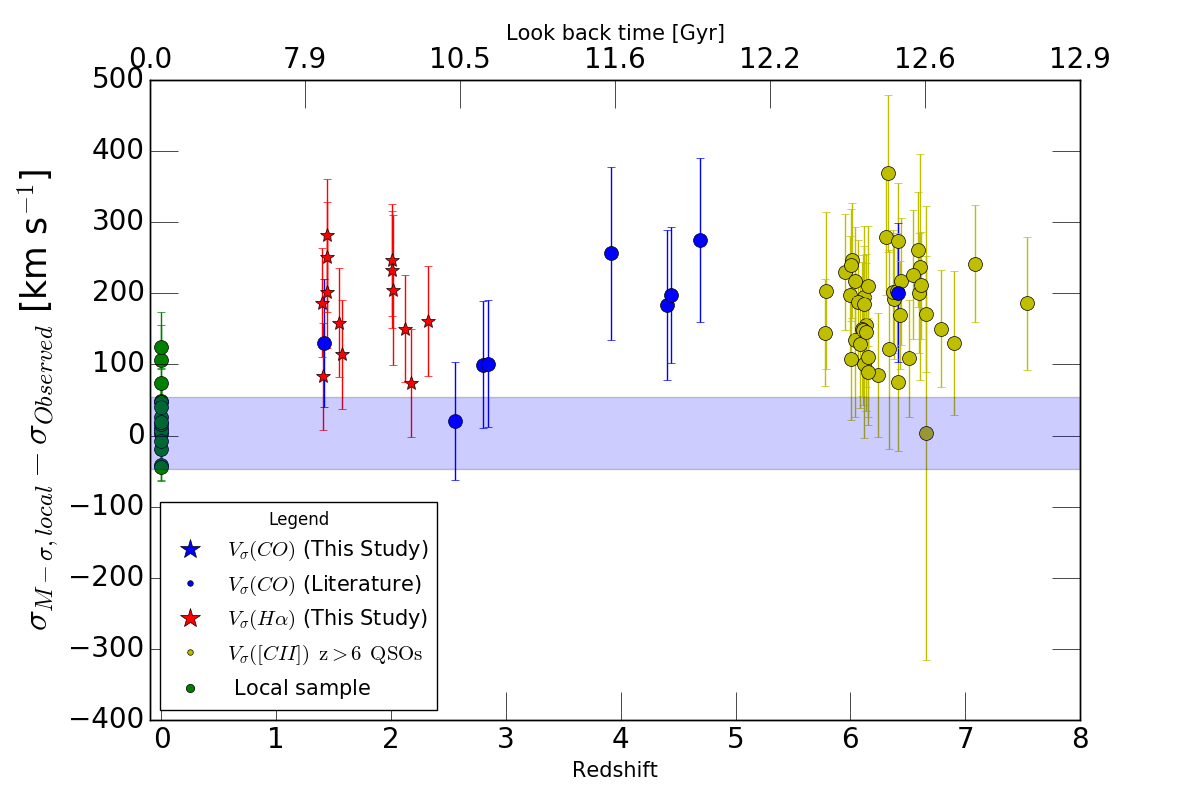}
    \caption{Measured offset of galaxies from the local \msigma scaling relationship (\cite{McConnell13}, $\log_{10}(M_{BH}/M_{\odot})=8.32+5.64\log_{10}(\sigma/200 \rm \enspace km \enspace s^{-1})$). On the y-axis, we quantify the offset as the difference between the observed and predicted velocity dispersion from the local scaling relation based on the observed SMBH mass. We plot the observed offset from the local scaling relation against the redshift for individual targets. The labels are similar to \ref{fig:m-sigma}. The shaded blue region represents the intrinsic scatter in the \msigma relationship for black holes with a mass of $10^{9.5}$ \msun. There is an overall offset for galaxies with massive SMBH at z$>$1 from the local \msigma relationship. We find no statistically significant difference in the offset between any of the high redshift samples, while there is a statistically significant offset from the local BCG points (green).}
    \label{fig:cosmic_m_sigma_offset}
\end{figure*}

In Figure \ref{fig:cosmic_m_sigma_offset}, we plot the offset from the local scaling relation against the redshift of each object from our sample, the local galaxies sample with SMBH $>10^{9}$ \msun and higher redshift quasars. Quasars with SMBH $>10^{9}$ \msun appear to be offset from the local scaling relationship, which indicates that SMBH growth appears to outpace that of stars in these systems. The SMBHs may grow rapidly up to a mass of several times $10^{9}$ \msun as early as 690 Myr after the Big Bang \citep{banados18}, matching in mass to some of the most massive SMBH seen today. Some galaxies with lower luminosity AGN and lower mass SMBH also appear to be offset from the local scaling relation at z$>1$ \citep{Merloni10, Bennert11}. Given the typically large uncertainty on the measured values and generally small sample sizes, it is difficult today to say whether a different population of AGN/galaxies are offset differently from the local scaling relationships at z$>1$.

Under the assumption that SMBH primarily grows through Eddington-limited gas accretion, the growth is expected to be exponential. The e-folding or ``Salpeter" time scale is about 50-300 Myr, depending on the SMBH spin. At the mean redshift of our sample (z=1.87), the SMBHs are expected to experience 30-200 e-folds in mass growth. However, for a duty cycle of around $10\%$ \citep{Wang06} the expected number of e-folds drops down to about 3-20. Furthermore, the quasars in our sample are not accreting near the Eddington limit and can eventually switch from high to low accretion-rate mode, further decreasing the Eddington ratio. Hence, the SMBHs in our sample have nearly finished forming and will only further grow by a factor of 1.2-7 under the assumption of an Eddington ratio of 10$\%$, and a duty cycle of 10$\%$. If these galaxies are to assemble onto the local scaling relation and to evolve into the most massive early-type galaxies that we see today, then the rapid SMBH growths at early times in the Universe must be followed by significant stellar growth. On average, the galaxies within our sample need to grow the stellar mass within a radius of 7 kpc at a constant rate of 100 \myr from z=2 to z=0. 

In the host galaxy of 3C 298, there is currently insufficient molecular gas for the galaxy to grow in stellar mass to match the mass predicted by the local scaling relationship. Furthermore, the quasar 3C 298 does not appears to live in an over-dense environment based on the number count of galaxies seen with the Spitzer space telescope imaging data \citep{Ghaffari17}. The open question is, how do these galaxies obtain the stellar mass necessary to grow into the massive galaxies we see today? Are minor mergers responsible for growing these galaxies? Alternatively, is the accretion of cool gas from the CGM responsible for providing the fuel necessary for future star formation?  The results we find for the host galaxy of 3C 298 favor the scenario where cold accretion flows from the CGM will supply most of the fuel necessary for future star formation. Another scenario could be that the Spitzer observations are too shallow to see lower mass galaxies. If these systems are gas-rich, they can supply future fuel for star formation from merging the gas in their CGM and ISM with the quasar's host. Indeed in recent hydrodynamical simulation \citep{Angles-Alcazar17} found that for dark matter halos with masses $>10^{12.5}$ \msun majority of the mass build up happens from gas accreted from the CGM and transfer/exchange of gas from CGM and ISM of cannibalized low mass galaxies. These simulations also find that stellar build-up from dry mergers and just accretion of stars from merging galaxies is not significant to grow the stellar mass of galaxies in massive halos. If this is the case for the majority of our galaxies, it implies that they have enormous amounts of gas inside their CGM. 

Our results can be in stark contrast to the predicted evolutionary paths of massive galaxies. In today's theoretical framework \citep{DiMatteo05, Hopkins08, Zubovas12, Zubovas14}, feedback from the SMBH is predicted to happen once the galaxy reaches the local \msigma relationship. However, our systems are showcasing outflows that are capable of causing major feedback when the mass of the galaxies is a fraction of their predicted final mass from the local scaling relations. Also, the gas-phase metallicities are far lower than those observed in nearby AGN. The kinetic luminosities for half of the outflows in our sample are far lower than the values predicted in simulations for the bolometric luminosities of our quasars \citep{vayner19a}. Ionized outflows in other samples show similar results, where about half the objects lie below the predicted minimum energy-coupling between the quasar and the outflow of 0.1$\%$ at z$\sim2$ \citep{vayner19a}. If all these systems are offset from the local scaling relationship, it would be easier to launch the outflows because their masses are smaller compared to if they were on the local scaling relations. This could lead to lower energy coupling efficiency. On the other hand, we might be missing a significant fraction of the gas within the outflows because a large portion of the gas could be in either a molecular or neutral phase.

In the quasar host galaxy of 3C298, we find the majority of the gas in the outflow is in a molecular state, and once combined with the ionized kinetic luminosity we find values closer to those predicted in simulations. The kinetic luminosity in 3C298 is close to 1$\%$ of the quasar's bolometric luminosity. Regardless if we are accounting all the gas in the outflow, outflows capable of causing feedback are occurring before the galaxies are on the \msigma relationship. We might need to reconsider our theoretical framework for massive galaxy formation, where the gas is not cleared from the galaxy in a single ``burst" of feedback once the galaxies reach the \msigma relationship. Instead, the SMBH grows first in massive dark matter haloes, followed by a delayed growth of the host galaxy with regulatory feedback from the SMBH and near-continuous accretion of gas from the CGM and nearby satellite galaxies. In such a scenario, the coupling efficiency might be lower per outflow event, compared to a single burst model where a single outflow-event clears all the gas. At later times, maintenance mode feedback from jets can heat the CGM, preventing gas from cooling and accreting onto the galaxy, keeping the galaxies ``quenched".

\section{Conclusions}\label{sec:conclusions}
We have conducted a near diffraction-limited survey of 11 quasar host galaxies to study the distribution, kinematics, and dynamics of the ionized ISM using the OSIRIS IFS at the W.M. Keck Observatory. This survey paper aimed to understand the source of gas ionization, the physical and chemical conditions of the ISM and to estimate the masses of radio-loud quasar host galaxies at z$\sim$2. We detected extended emission in all objects on scales from 1-30 kpc and found that:

\begin{itemize}

\item The AGN photoionizes the majority of the extended gas. A significant fraction of emission-line ratios are found to reside between the two sequences on the traditional BPT diagram. By applying evolutionary models of the mixing and star-forming sequence from z=0 to z=2.5, we find that the gas within our systems is denser and has lower metallicity compared to the gas photoionized in local AGN.

\item In 9 objects, we find dynamically quiescent regions, with lower average \logohb ratios. For systems where Hubble Space Telescope imaging is available, their morphologies are consistent with clumpy star-forming regions commonly observed in the distant Universe, indicating the presence of recent star formation. We find these systems to be forming stars at a maximum rate of 9-160 \myr based on the \ha luminosity.

\item For nine objects, we are able to measure the mass of the SMBH, the stellar velocity dispersion using the narrow component of \ha emission line as a proxy, and galaxy mass. We compare these nine objects to the local scaling relation between the mass of the SMBH and the mass or velocity dispersion of the galaxy. Our systems are both offset from the \msigma and \mstellar relationship. Substantial growth is still necessary if these systems are to evolve into the present-day massive elliptical galaxies. Gas accretion from the CGM and gas-rich minor mergers are necessary to grow the stellar mass and increase the metallicity of the ISM. On average, the galaxies need to grow by at least an order of magnitude in stellar mass if they are to assemble onto the local scaling relations. A near-constant mass growth rate of $\sim$100 \myr is necessary within a radius of 10 kpc from the quasar from z$\sim2$ to 0.

\item Combining the results of this paper with \citep{vayner19a} we find evidence for the onset of feedback before the galaxies are on the local \msigma relationship. Luminous type-1 quasars are not the end phase of massive galaxy formation. 

\end{itemize}

\acknowledgments
The authors wish to thanks Jim Lyke, Randy Campbell, and other SAs with their assistance at the telescope to acquire the Keck OSIRIS data sets. We want to thank the anonymous referee for their constructive comments that helped improve the manuscript. The data presented herein were obtained at the W.M. Keck Observatory, which is operated as a scientific partnership among the California Institute of Technology, the University of California and the National Aeronautics and Space Administration. The Observatory was made possible by the generous financial support of the W.M. Keck Foundation. The authors wish to recognize and acknowledge the very significant cultural role and reverence that the summit of Maunakea has always had within the indigenous Hawaiian community. We are most fortunate to have the opportunity to conduct observations from this mountain. This research has made use of the NASA/IPAC Extragalactic Database (NED) which is operated by the Jet Propulsion Laboratory, California Institute of Technology, under contract with the National Aeronautics and Space Administration.

\software{OSIRIS DRP \citep{OSIRIS_DRP}, Matplotlib \citep{Hunter07}, SciPy \citep{2020SciPy-NMeth}, NumPy \citep{2020NumPy-Array}, Astropy \citep{Astropy18}, MAPPINGS \citep{Alarie19}, emcee \citep{Foreman-Mackey13}}

\appendix\label{sec:appendix}

\begin{figure*}
    \centering
    \includegraphics[width=7in]{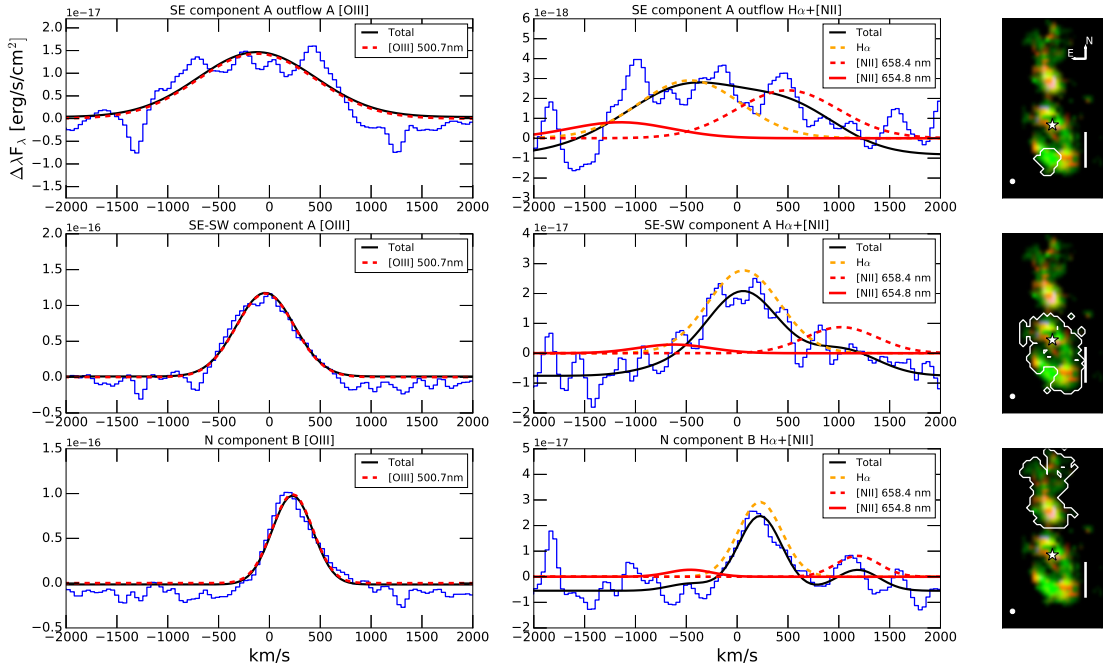}
    \caption{Spectra of distinct regions along with fits to individual emission lines for the 3C 9 system.}
    \label{fig:3C9_all}
\end{figure*}

\begin{figure*}
    \centering
    \includegraphics[width=7in]{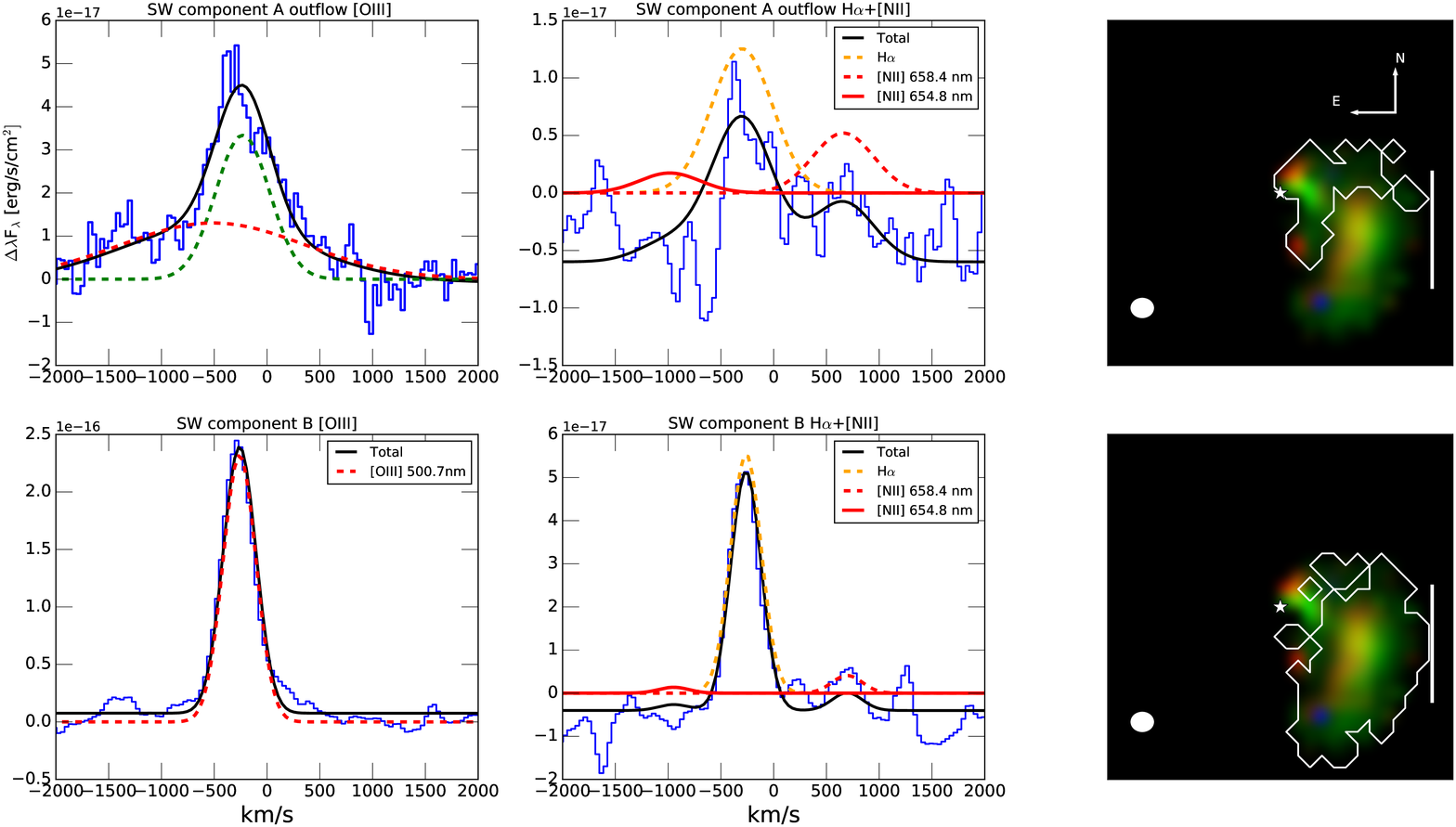}
    \caption{Spectra of distinct regions along with fits to individual emission lines for the 3C268.4 system.}
    \label{fig:3C2684_all}
\end{figure*}

\begin{figure*}
    \centering
    \includegraphics[width=7in]{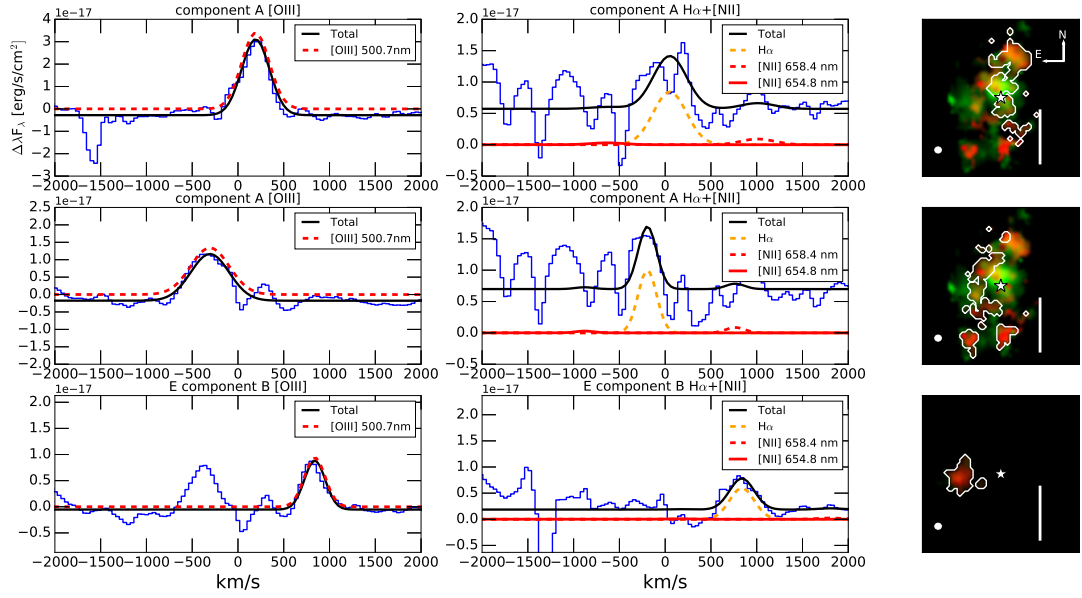}
    \caption{Spectra of distinct regions along with fits to individual emission lines for the 7C 1354+2552 system.}
    \label{fig:7C1354_all}
\end{figure*}

\begin{figure*}
    \centering
    \includegraphics[width=7in]{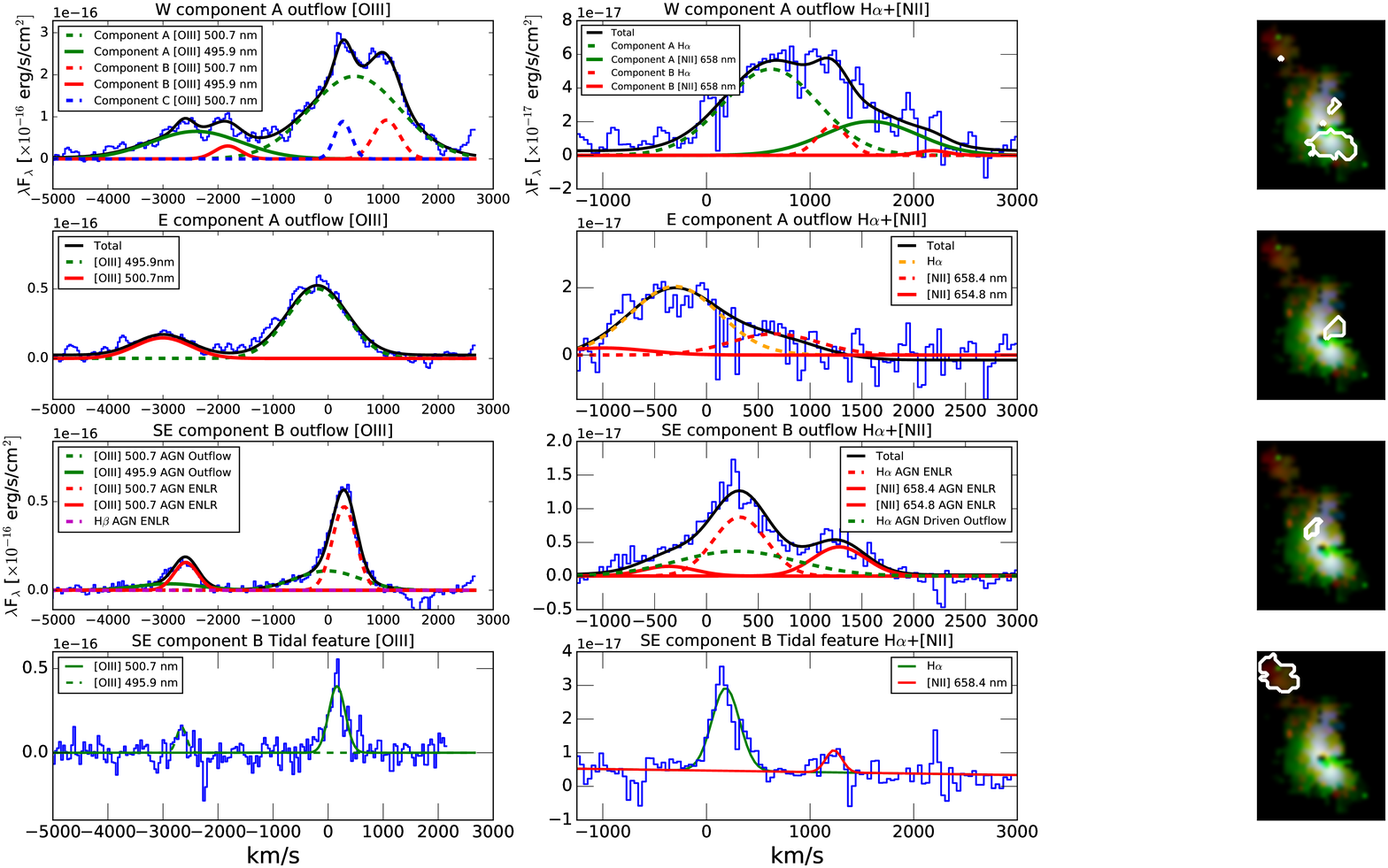}
    \caption{Spectra of distinct regions along with fits to individual emission lines for the 3C 298 system.}
    \label{fig:3C298_all}
\end{figure*}

\begin{figure*}
    \centering
    \includegraphics[width=7in]{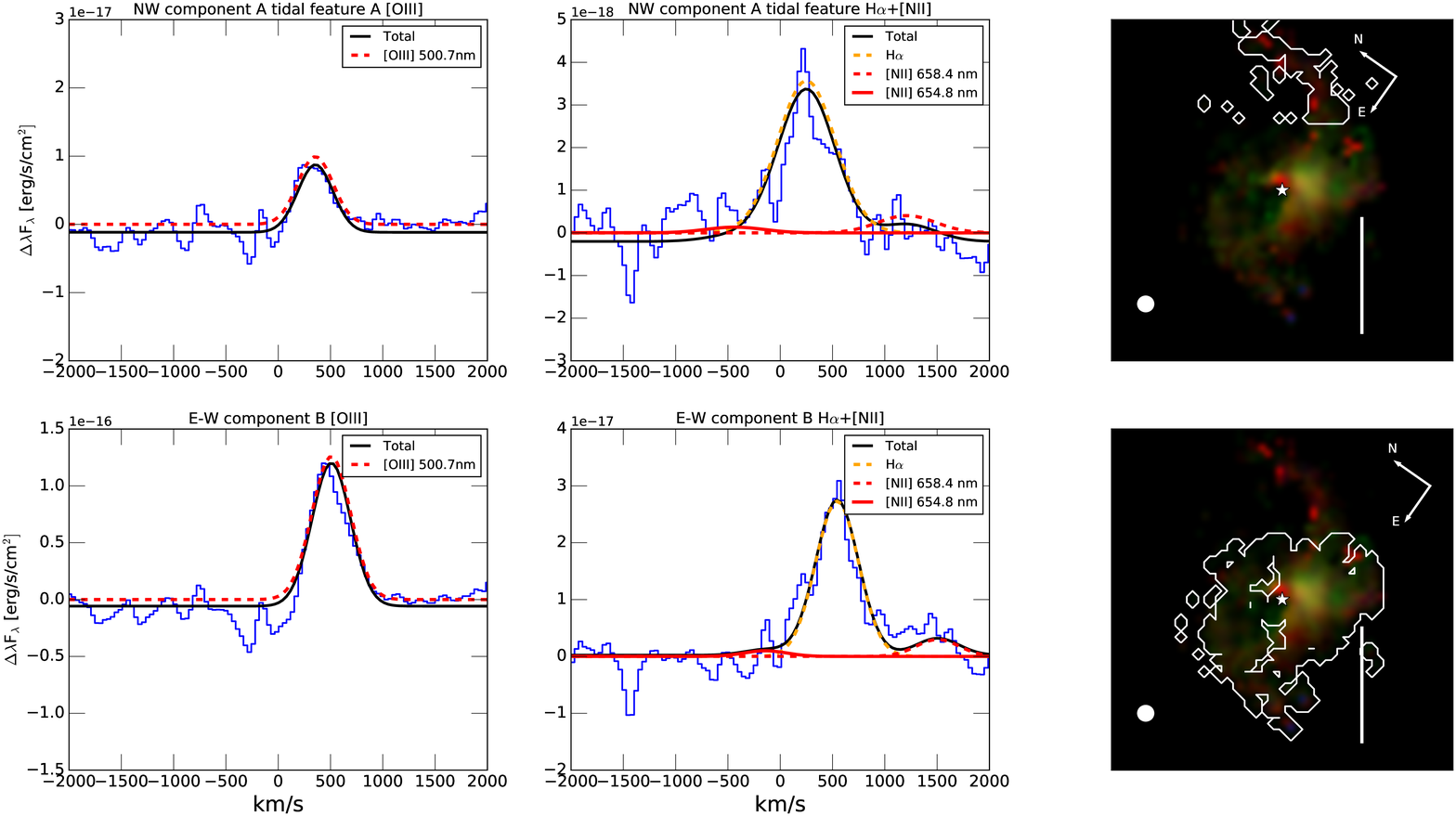}
    \caption{Spectra of distinct regions along with fits to individual emission lines for the 3C 446 system.}
    \label{fig:3C446_all}
\end{figure*}

\begin{figure*}
    \centering
    \includegraphics[width=7in]{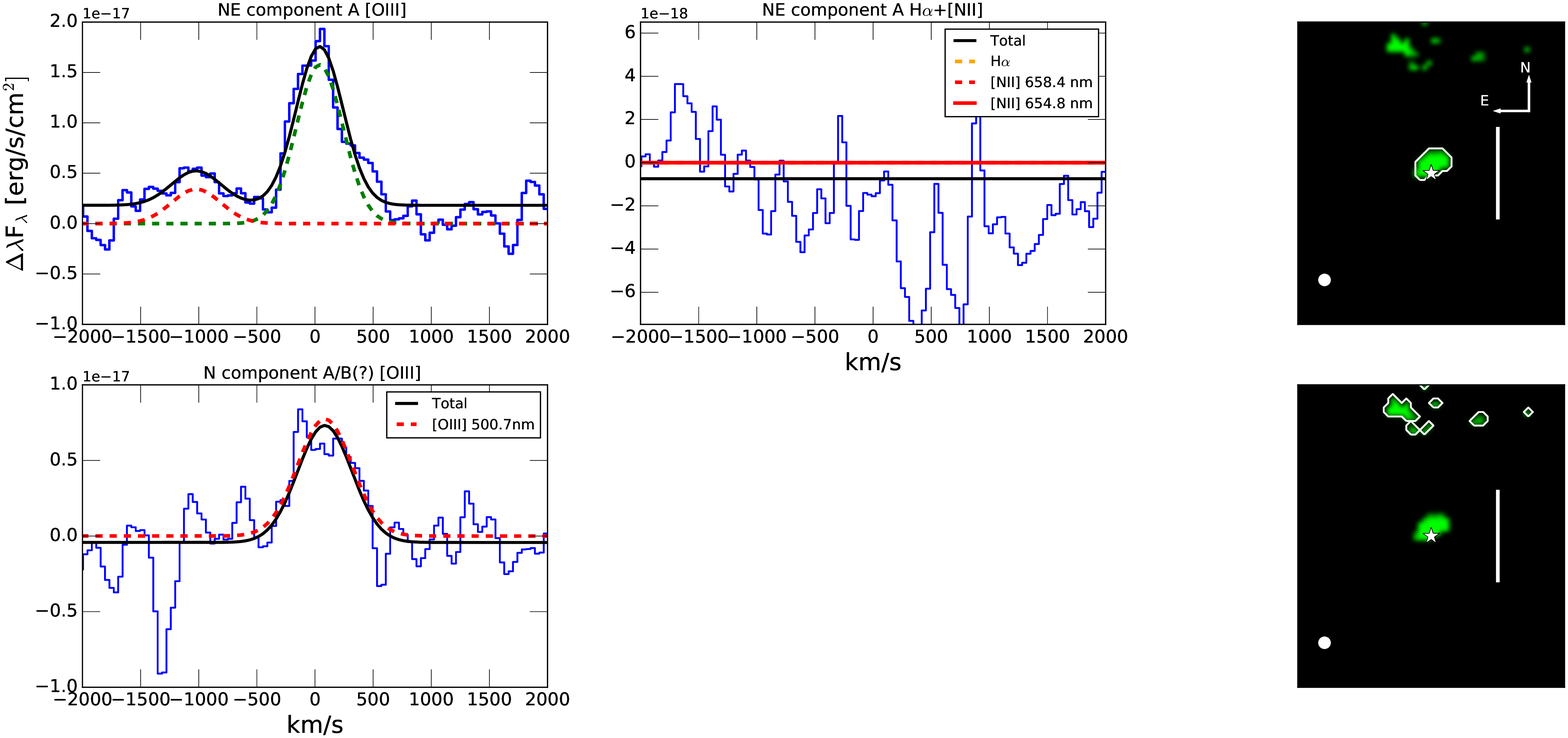}
    \caption{Spectra of distinct regions along with fits to individual emission lines for the 4C 57.29 system.}
    \label{fig:4C5729_all}
\end{figure*}

\begin{figure*}
    \centering
    \includegraphics[width=7in]{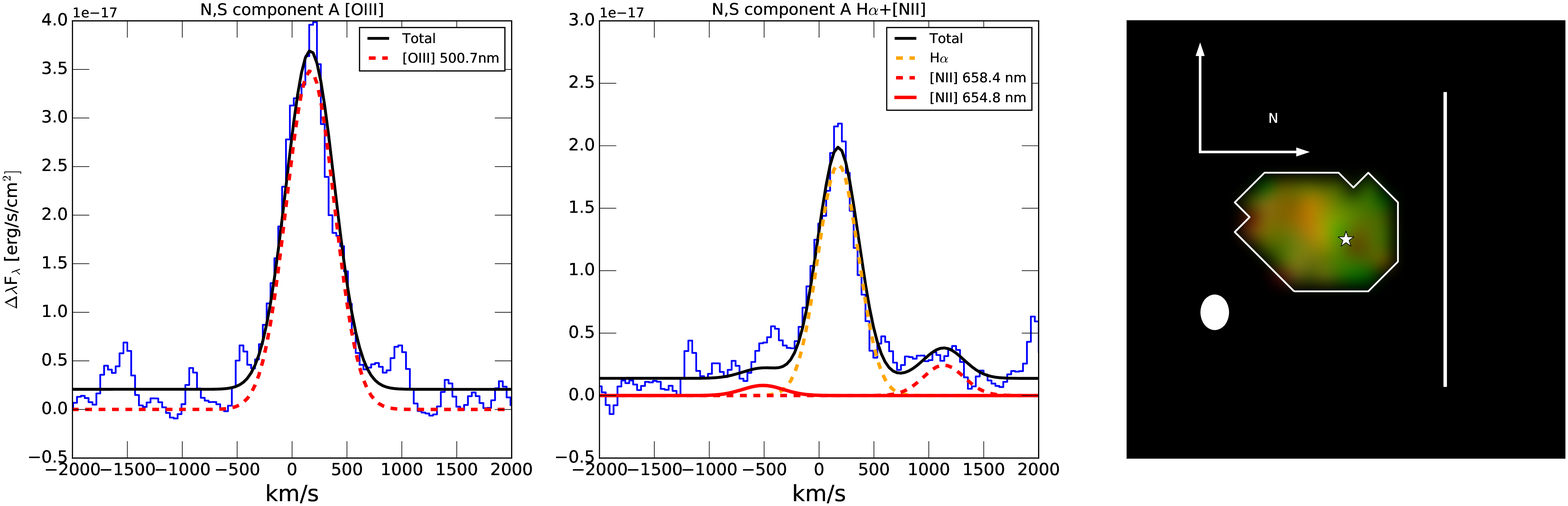}
    \caption{Spectra of distinct regions along with fits to individual emission lines for the 4C 22.44 system.}
    \label{fig:4C2244_all}
\end{figure*}

\begin{figure*}
    \centering
    \includegraphics[width=7in]{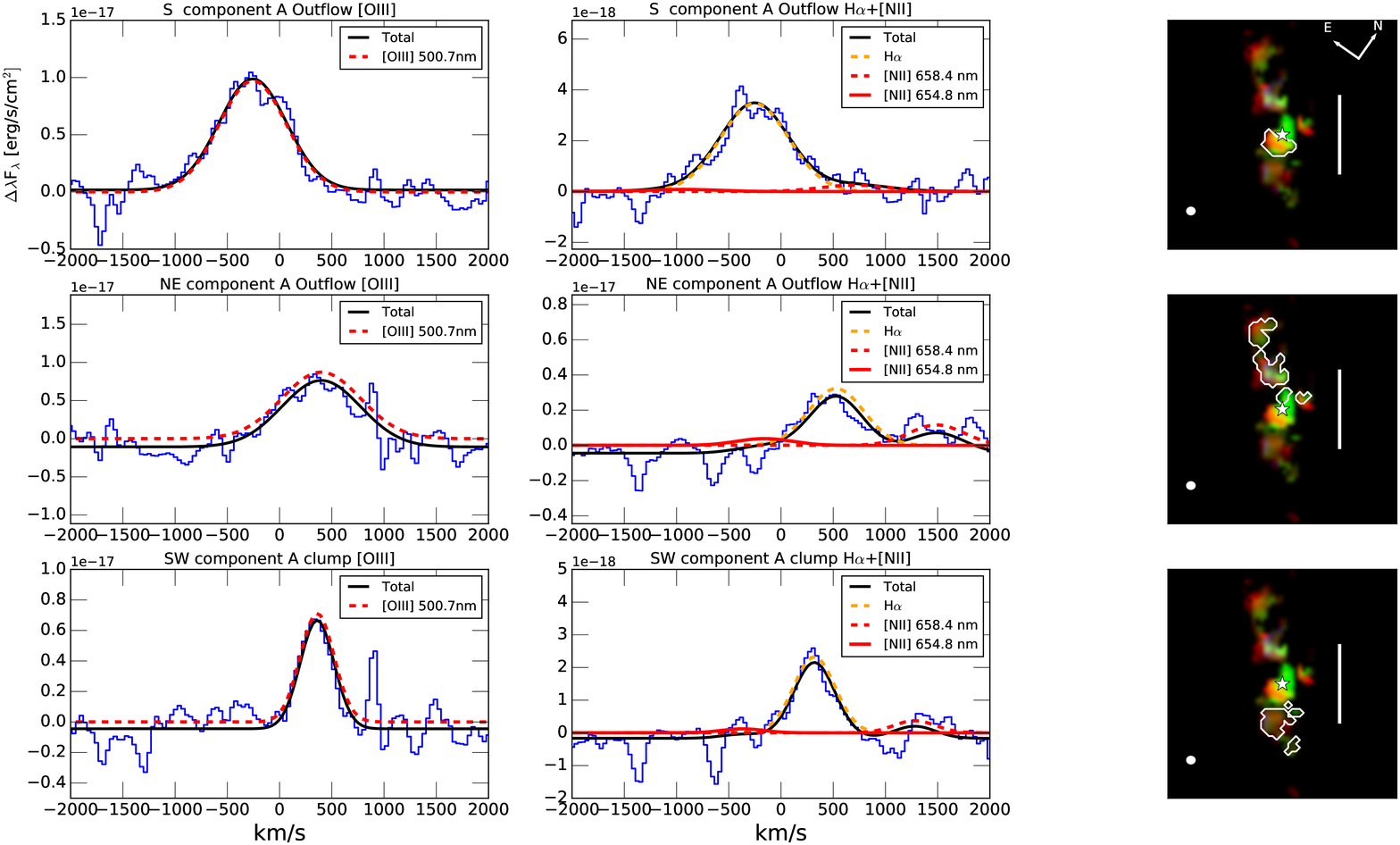}
    \caption{Spectra of distinct regions along with fits to individual emission lines for the 4C 05.84 system.}
    \label{fig:4C0584_all}
\end{figure*}
\section{3C 9}\label{app:3C9}
3C9 is a luminous quasar at z = 2.019922 with a prominent blue rest-frame UV continuum. For this source, we identify three distinct regions. ``SE-SW component A" is a region with a ring-like morphology associated with the 3C9 quasar host galaxy. We measure a velocity dispersion from a Gaussian fit to the nebular emission lines of 407.6$\pm$12.9\kms and the kinematics resembling a rotating disk. ``SE component A" is classified as an outflow region with a very high emission line FWHM of 1362.7$\pm$60.5 \kms and elevated \logohb and \loghn~ratios relative to the rest of the system. ``N component B" is the merging galaxy in the 3C9 system showcasing a line FWHM of 472.15$\pm$11.8 \kms and a velocity offset of $\sim$200 \kms from the quasar. The projected spatial separation between the two apparent nuclei is 9 kpc. The quasar lies in the galaxy with a ring-like morphology showing the kinematic structure of a disk. Archival HST imaging of rest-frame UV continuum shows the ring morphology as well (see Figure \ref{fig:quiescent-hst}), indicating very recent star formation activity in the ring. The merging galaxy ``N component B" appears to be a dispersion dominated system with active star formation and appears in rest-frame UV emission. The 3C9 system best resembles the local galaxy merger system Arp 148 (z=0.036), also known as Mayall's Object. The outflow in this system appears to be emanating from the ring of the galaxy with the quasar.

\section{4C 09.17} \label{app:4C09.17}

4C 09.17 is a luminous quasar at z=2.117 with a blue UV continuum. For this source, we identify four distinct regions. ``SW component A" is a star-forming clump associated with the quasar host galaxy. The spectrum of this region shows a single narrow emission line with an FWHM of 312.0$\pm$7 \kms. ``S/E component A" is an outflow region driven by the quasar, the nebular emission lines for this region have an FWHM of 887.2$\pm$22.4 \kms. A second narrow component is required for a good fit for each emission line in this region, with a line FWHM of 290.4$\pm$29.9 \kms. ``W component B clumps" is a region part of the merging galaxy within the 4C09.17 system. The region consists of clumpy emission selected by isolating spaxels with an \ha line surface density $>6\times10^{-16}$ \ferg arcsec$^{-2}$. ``W component B diffuse" is emission associated with ``diffuse" ionized emission in the merging galaxy selected by isolating spaxels with an \ha spatial line surface density $<6\times10^{-16}$ \ferg arcsec$^{-2}$. The diffuse region shows higher \logohb and \loghn~line ratios associated with both AGN and star formation photoionization while the clumpy regions of the merging galaxy showcase lower ionization levels consistent with photoionization by star formation. This region is associated with bright UV emission in HST imaging of this object \citep{Lehnert99}. ``S/E component A outflow" shows high \loghn~and \logohb values relative to the rest of the system, indicating this region is predominantly photoionized by the quasar. The 4C09.17 system is a merger of two galaxies with velocity offsets of $\sim$1000 \kms and a projected separation of $\sim$ 4 kpc. HST imaging of rest-frame UV continuum (see Figure \ref{fig:quiescent-hst}) shows evidence for a population of young hot stars indicating recent star formation activity. The majority of the star formation activity is confined to the merging galaxy.

\section{3C 268.4} \label{app:3C268.4}
3C 268.4 is a luminous quasar at z=1.39, with a slightly reddened UV continuum compared to the average type-1 quasar. For this target, we identified two distinct regions. ``SW component A" is an outflow driven by the quasar. The FWHM of the emission lines is 2075$\pm$354 \kms as measured from the Gaussian fit to the \oiii line. The spectrum extracted over this region also shows a narrow component with an FWHM of 603.7$\pm$54.9 \kms, most likely signaling emission from an extended narrow-line region close to the quasar. Because of issues with miss-assignment of flux in the OSIRIS pipeline \citep{Lockhart19}, the rows below and above the centroid of the quasar do not have properly extracted spectra in the H band observations of this object. Hence we do not have a good spectrum of the extended emission in a 0.2-0.3\arcsec~radius around the quasar in the H band, which covers the \ha and \nii emission lines of the ionized outflow. ``SW component B" is a region associated with the merging galaxy, showcasing clumpy morphology in ionized gas emission. The emission lines have an FWHM of 367.7 $\pm$ 3.9 \kms and an offset of $-300$ \kms relative to the redshift of the quasar. The \logohb line ratios are lower for this region compared to the rest of the system, consistent with a mixture of AGN and star formation photoionization. This region is also associated with bright rest-frame UV continuum emission, seen with HST observations of this target \cite{Hilbert16}.

\section{7C 1354+2552} \label{app:7C1354+2552}
7C 1354+2552 is a luminous quasar at z=2.0064 with a blue UV continuum. For this target, we identify two distinct regions. ``Component A" is the extended emission associated with the quasar host galaxy. The kinematics show a smooth velocity gradient, indicating the presence of a galactic disk. The size, morphology, and kinematics of the disk are similar to that of star-forming galaxies on the more massive end of the star formation main sequence at $z\sim$2 \citep{Forster18}. We measure an emission line FWHM of 357.2$\pm$2.0 \kms on the redshifted side of the disk and 497.7$\pm$6.5 \kms on the blue-shifted side of the disk. Although this region only has a single label (``component A"), in Figure \ref{fig:7C1354_all} rows one and two show the fits to the red and blue-shifted sides of the disk that are part of this region. This region is selected based on the location where \ha emission is detected. This is done to boost the SNR in the \ha line as it appears to be clumpier, more compact, and less extended than \oiii. In Table \ref{tab:fluxes} we provide values integrated over the entire galactic disk. ``E component B " is a region associated with the merging galaxy at a projected separation of 6-7 kpc. The kinematics are consistent with a dispersion dominated galaxy. The entire ``component A" is consistent with quasar photoionization. The gas in ``E component B" is photoionized by star formation. 

\section{3C 298}\label{app:3C298}
3C298 is a luminous quasar at z=1.439 with a slightly reddened UV continuum. For this target, we identify five distinct regions. We present a detailed discussion of each region in \cite{Vayner17}. ``W/E component A" are outflow regions with a bi-conical morphology, where the western (W) is the redshifted receding cone, and the eastern (E) is the approaching cone. In \cite{Vayner17}, they are referred to as the red(blue) shifted outflow region. The emission lines over the outflows are very broad, with FWHM up to $\sim$1500 \kms. A combination of shocks and quasar activity is likely responsible for photoionizing the gas. ``SE component B outflow" is an outflow region belonging to a merging galaxy. ``SE component B ENLR" is an extended narrow-line region belonging to the disk of the merging galaxy, with gas photoionized by the quasar or secondary AGN. ``SE component B Tidal feature"  is a region of the merging galaxy with active/recent star formation as evident by lower \loghn~and \logohb values compared to the rest of the regions.

\section{3C318}\label{app:3C318}

3C318 is a luminous quasar at z=1.5723 with a reddened UV continuum. There is evidence for a spatially unresolved nuclear star-burst with an upper limit on the star formation rate of 88$\pm$9\myr. This star formation rate is far lower than the far infrared derived rate of 580\myr. The extinction towards the nuclear region measured from \cite{Willott00} alone cannot explain the mismatch between the \ha and far-infrared derived SFR. Either a larger fraction of the far-infrared emission is from dust that is being heated by the AGN, or the far-infrared emission traces a different star formation history than \ha \citep{Calzetti13}. No narrow extended emission is detected in this object.

The merger status of this object is unclear. Two nearby galaxies to the north and west of the quasar are visible in archival HST imaging \citep{Willott00}. We do not detect the western object that is 2\arcsec~away from the quasar in our OSIRIS observations in any emission line. \cite{willott07} studied this object with PdBI through CO 2-1 emission at a fairly coarse ($\sim$8 arcseconds) resolution. There appears to be CO emission that could plausibly be associated with the western object. We have recently obtained a much higher angular resolution CO 3-2 spectroscopy of this target that will be discussed in detail in a forthcoming paper. We confirm the existence of CO 3-2 emission associated with the CO 2-1 emission. We resolve the molecular emission into multiple components. However, the CO 3-2 emission is not associated with either one of the galaxies seen in the HST data. We obtained a wide field of view IFS observations of this target with KCWI aimed at attempting to measure the redshifts of the nearby galaxies and to confirm the merger scenario of this object. We detect both the northern and western objects in the continuum. We confirm that the northern target is at a different redshift than the quasar from the detection of [OII] emission, while for the western object, a reliable redshift is challenging to determine with the current data set. Hence no clear evidence of a companion galaxy that is part of a merger is detected for this quasar associated with the brightest galaxies seen in optical imaging within a few arcseconds from the quasar.

\section{4C 57.29}\label{app:4C57.29}
4C 57.29 is a luminous quasar at z=2.1759 with a blue UV continuum. For this target, we identify two regions. Region ``NE component A" belongs to the host galaxy of the quasar. The relatively high \logohb value indicates that this region is consistent with being photoionized by the quasar. The 500.7 nm \oiii is the only emission line detected for this region. The region is marginally resolved, making it hard to measure the kinematic structure. We require a double Gaussian fit to the \oiii emission in this region to obtain a good fit, and we measure an FWHM of 474.3 and 502.5 \kms with offsets of 35.0 \kms and -1050.1 \kms relative to the redshift of the quasar. We identify a second region north of the quasar. It is unclear if it belongs to a merging galaxy or the quasar host galaxy. There is a $\sim100$ \kms offset from the quasar, and the line has an FWHM of 550.13$\sim$19 \kms. This region is also only detected in \oiii. The SNR is too low to measure any kinematics structure. 

\section{4C 22.44}\label{app:4C22.44}
4C22.44 is a luminous quasar at z=1.5492 with a reddened UV continuum. Similar to 3C318, we do not detect any evidence for a merging galaxy for this system. For this target, we identify a single region, ``N, S component A". The kinematics of this region may be consistent with a galactic disk belonging to the quasar host galaxy. We see evidence for a smooth gradient in the radial velocity map, however, the region is marginally resolved. We measure an emission line FWHM of 434.8 \kms. The region is consistent with being ionized by star formation with some contribution from quasar photoionization. 

\section{4C 05.84}\label{app:4C05.84}
4C05.84 is a luminous quasar at z=2.323 with a slightly reddened UV continuum. For this target, we identify three distinct regions. Regions ``S component A" and ``NE component A" are the blue(red) shifted outflow regions resembling a bi-conical outflow. They showcase broad extended emission with a line FWHM of $\sim$800 \kms. The quasar photoionizes these regions. Region ``SW component A clump", shows a line FWHM of 467.9$\pm$3.0 \kms and is photoionized by a combination of star formation and the quasar. This clump is also detected in NIRC2 imaging of this object studied by \cite{Krogager16}, where they consider this clump to be associated with a damped Ly$\alpha$ system. However, here we confirm that this objected is part of the quasar host galaxy. We find no evidence for a merging galaxy within our OSIRIS observations.

\section{3C 446}\label{app:3C446}
3C446 is a quasar at z = 1.404. For this target, we identify two regions, ``N component A tidal feature" is a region belonging to the quasar host galaxy, resembling a tidal feature that is most likely induced by the merger. We measure an FWHM of 395.14$\pm$2.0 \kms for this region. ``E-W component B" belongs to the merging galaxy, a portion of it resembles a tidal feature, counter to the tidal arm of ``N component A tidal feature." For this region, we measure a line FWHM of 558.5$\pm$63 \kms however, it appears to be a blend of two velocity components. It is unclear where the nucleus of the merging galaxies resides. It could be that it has already merged with that of the quasar host galaxy. The two galaxies appear to be offset by at least 500 \kms in velocity, and there is a possibility that a portion of the merging galaxy lies on top of the quasar host galaxy.

\section{4C 04.81}\label{app:4C0481}
4C04.81 is a luminous quasar at z=2.5883 with a reddened UV continuum. For this target, we identify a single region, ``E component A outflow". The kinematics show blue and red-shifted broad (FWHM$\sim$800 \kms) emission. The quasar mainly photoionizes the gas. We do not identify any narrow extended emission in this object.


\bibliography{bib}
\end{document}